\documentclass{cernyrep}
\usepackage{texnames}
\usepackage[T1]{fontenc}
\usepackage{graphicx}
\usepackage{amsmath}
\usepackage{mathtools}
\usepackage{wrapfig}
\usepackage{multirow}

\usepackage{ragged2e}

\usepackage{tikz}
\usetikzlibrary{trees,arrows}
\usetikzlibrary{decorations.pathmorphing}
\usetikzlibrary{decorations.markings}

\tikzset{
  >=stealth',
every node/.style={font=\scriptsize},
   photon/.style={decorate, decoration={snake,amplitude=2pt,segment
        length=8pt,pre length=0cm,post length=0}},
    photontiny/.style={decorate, decoration={snake,amplitude=1pt,segment
       length=4pt,pre length=0cm,post length=0}},
    electron/.style={postaction={decorate},
        decoration={markings,mark=at position .55 with {\arrow{>}}}},
    antielectron/.style={postaction={decorate},
        decoration={markings,mark=at position .52 with {\arrow{<}}}},
    drelectron/.style={line width=1.5pt,postaction={decorate},
        decoration={markings,mark=at position .52 with {\arrow{>}}}},
    gluon/.style={decorate,
        decoration={coil,amplitude=4pt, segment length=5pt}},
    scalar/.style={densely dashed},
     cut/.pic ={
      \draw[line width=0.75pt] +(-135:.2cm) -- +(45:.2cm)
                               +(-45:.2cm) -- +(135:.2cm);}
}

\title{Flavor physics and CP violation}

\author{M.I. Vysotsky}
\institute{ITEP, Moscow, Russia}

\begin{document}

\maketitle

   OUTLINE

  \begin{enumerate}
  \item Introduction: Why $N_q = N_l$ and why we are sure that
    $N_g = 3$.

  \item Cabibbo-Kobayashi-Maskawa (CKM) matrix, unitarity triangles.

  \item CP, CP violation.

  \item $M^0 - \bar M^0$ mixing, CPV in mixing.

  \item Neutral kaons: mixing ($\Delta m_{LS}$) and CPV in mixing
    ($\tilde\varepsilon$).

  \item Direct CPV.

\item Constraints on the unitarity triangle.

\item CPV in $B^0 - \bar B^0$ mixing.

  \item CPV in interference of mixing and decays,
    $B^0(\bar B^0) \to J/\Psi K$, angle $\beta$.

  \item What is the probability of $ \Upsilon(4S) \rightarrow B_d^0\bar B_d^0
\to J/\Psi K_{S}  \;J/\Psi K_{S}$ decay?

  \item CPV in $b \to s g \to s s
\bar{s}$ transition: penguin domination.

  \item $B_s(\bar B_s) \to J/\Psi \phi, \phi_s$.

  \item Angles $\alpha$ and $\gamma$.

  \item CKM fit.

  \item Perspectives: $K\longrightarrow\pi\nu\nu$, Belle II, LHC.

  \end{enumerate}

\newpage
\section{Introduction}
\subsection{Fundamental particles and Periodic Table}

  \vspace{-4mm}

  \begin{center}
    \begin{figure}[h]

     \includegraphics[width=.5\textwidth]{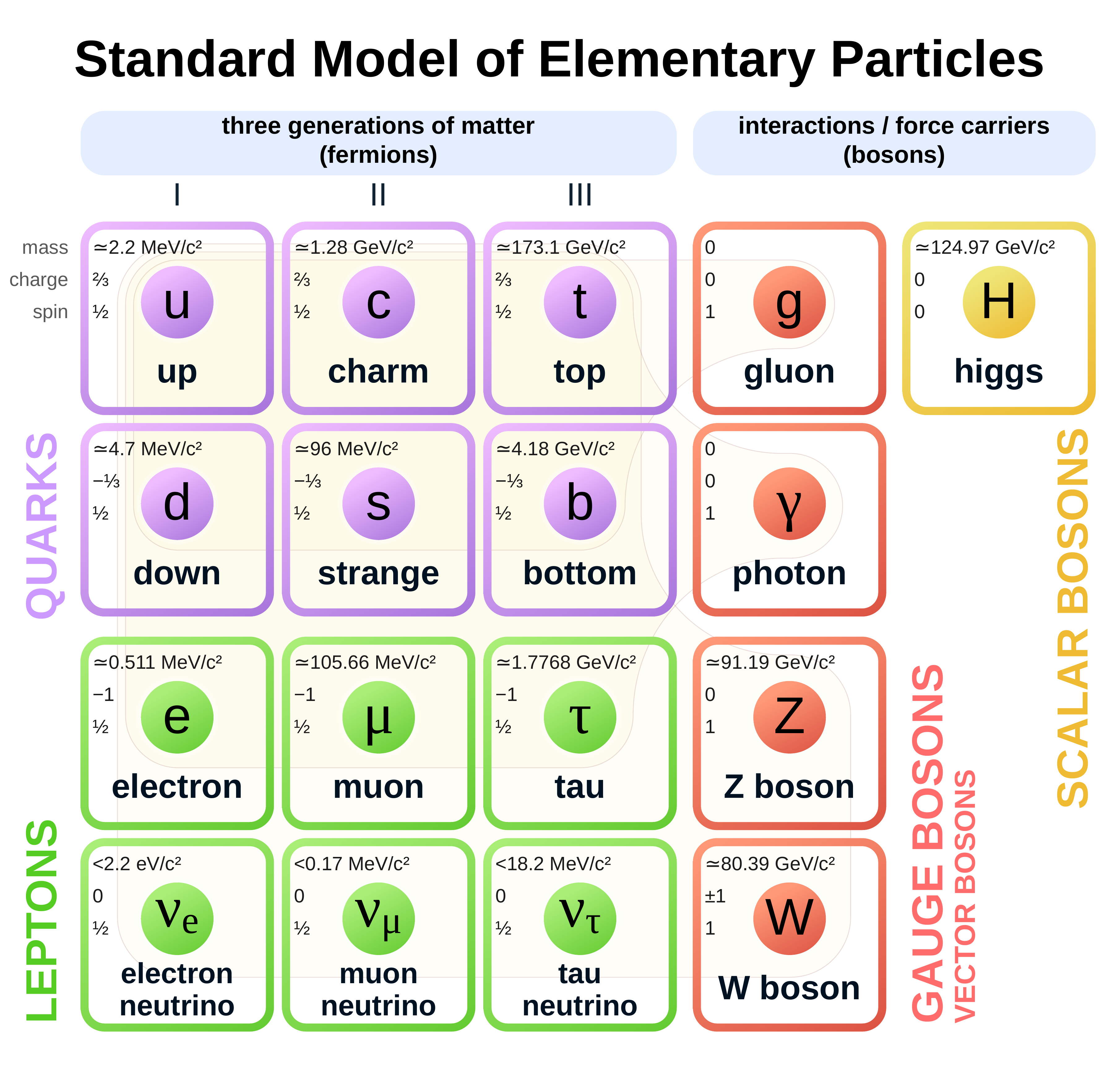}
    \end{figure}
  \end{center}

  \vspace{-8mm}

    One of the main problems for particle physics in the 21st century
    is why there are 3 quark-lepton generations and what explains fermion properties.
  This is a modern version of I.Rabi question which he asked in response to the news that a recently discovered muon is not a hadron: ``\textit{Who ordered that?}''\\

  \begin{center}
    \begin{figure}[h]

      \includegraphics[width=.7\textwidth]{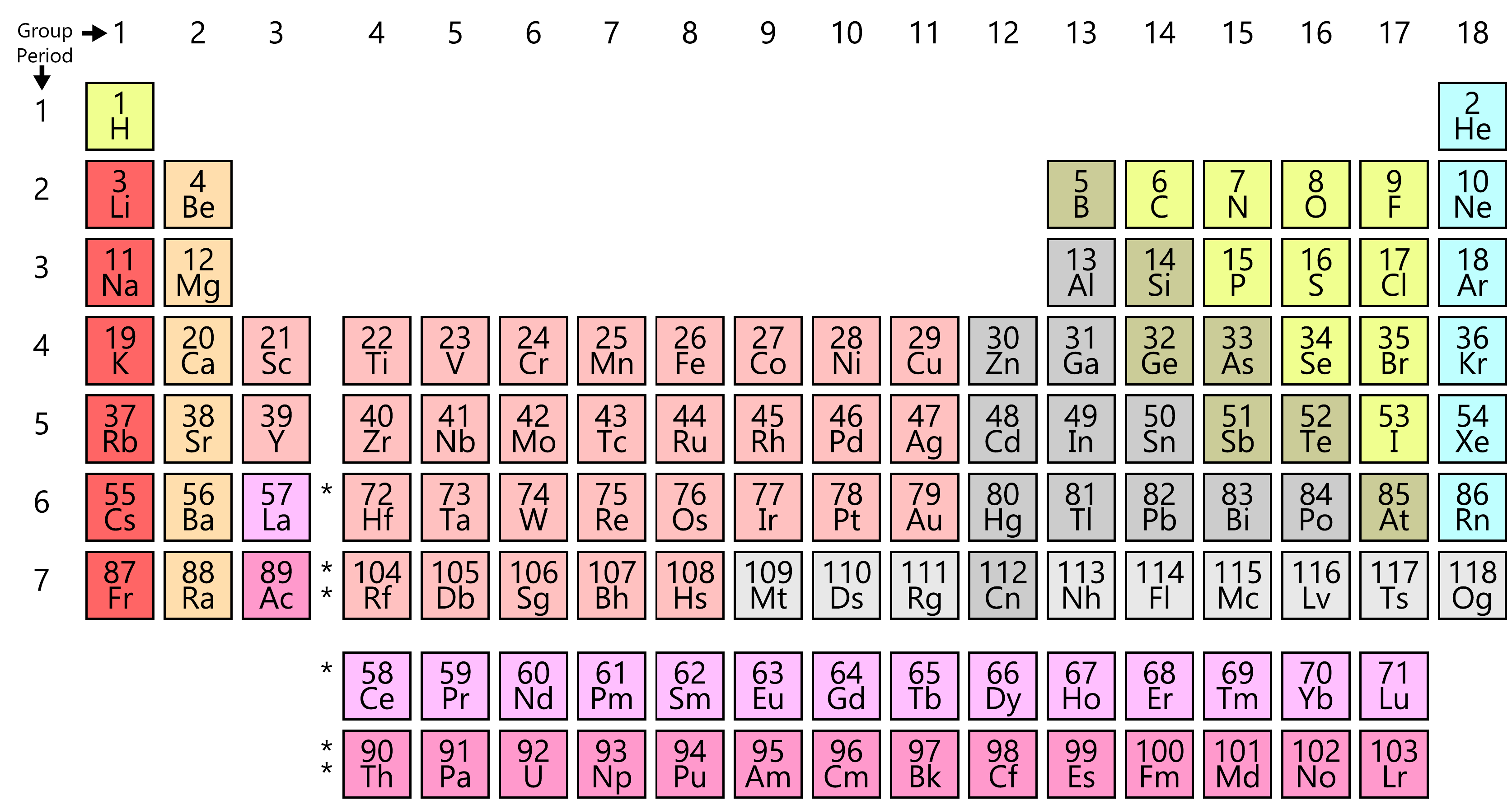}

    \end{figure}

  \end{center}
  Dmitry Mendeleev, professor of
 St. Petersburg University, discovered his Periodic Table in 1869, just 150 years ago.
 He put there 63 existing elements and predicted 4
new elements. This 19th century
  discovery was explained by Quantum Mechanics in the beginning of the 20th
  century. Let us hope that an explanation of the Table of Elementary
  Particles in general and the solution of a flavor problem in particular will be found
  in this century. There is much  in common with the  Periodic Table: $W,Z, H$ with their masses were predicted as well. The central question is: what is an analog of Quantum Mechanics?

\subsection{More generations?}

After the discovery of the third generation the speculations on the 4th generation were very
        popular. Why only 3?

  However for invisible Z boson width we have:

   \vspace{0.5cm}
  \begin{center}
    \includegraphics[width=.5\textwidth]{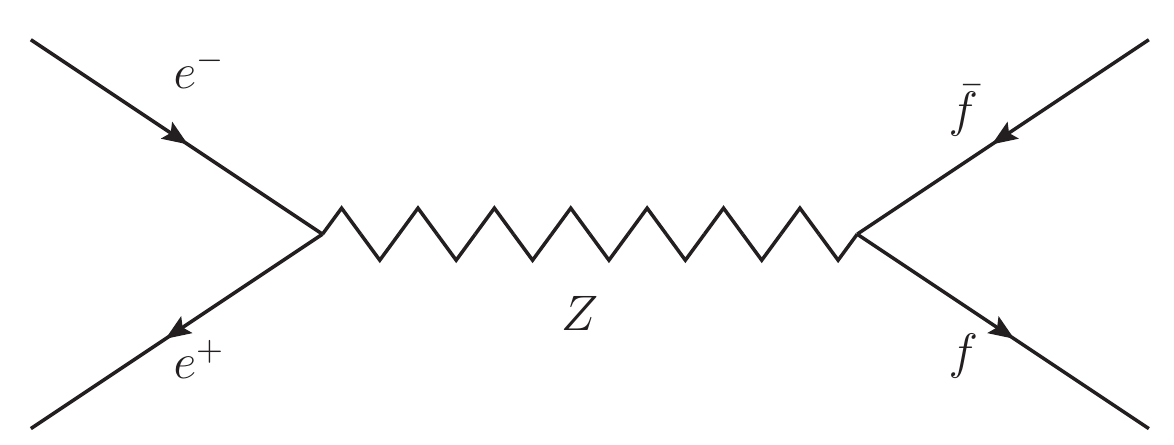}
  \end{center}

  \vspace{0.5cm}
  \begin{equation}
    \Gamma_{Z\to ff} = \frac{G_F M_Z^3}{6\sqrt{2}\pi} [(g_V^f)^2 +(g_A^f)^2] =
    332[(g_V^f)^2 +(g_A^f)^2] \; \mbox {\rm MeV}. \;\;
  \label{1}
  \end{equation}

  And taking into account $\nu_e, \nu_{\mu}$ and $\nu_{\tau}$ we obtain:
  \begin{equation}
    \Gamma_{Z\to \nu\nu}^{\mbox{theor}} = 3 \cdot 332[\frac{1}{4} +\frac{1}{4}] =
    498 \; \mbox{\rm MeV} \;\; ,
  \label{2}
  \end{equation}

  \begin{equation}
    \Gamma_{inv}^{\mbox{exp}} = 499 \pm 1.5 \; \mbox{\rm MeV} \;\; .
  \label{3}
  \end{equation}
  Thus $\nu_4$ is not allowed - so, there is no 4th generation.

  \bigskip
   BUT: what if $m(\nu_4)>M_Z/2$?

In H production at LHC the following diagram dominates:

\bigskip

\begin{center}
    \includegraphics[width=.5\textwidth]{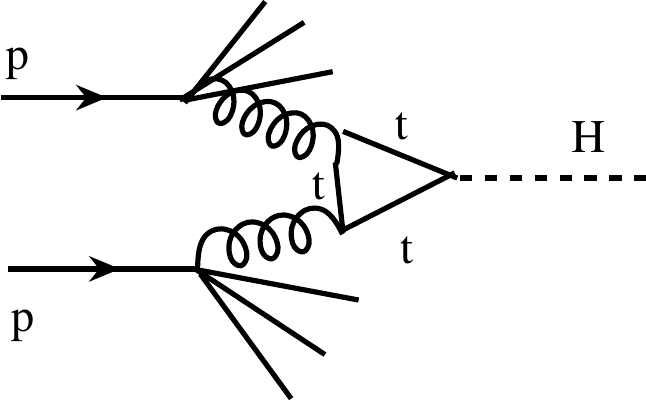}
  \end{center}
\bigskip

and for $2 m_t >> M_H$ the corresponding amplitude does not depend on  $m_t$.

  In case of the 4th generation $T-$ and $B-$ quarks
  contribute as well, so the amplitude triples and the cross section of $H$
  production at LHC becomes 9 times larger than in SM,
  which is definitely excluded by experimental data.

\bigskip
  Problem 1

     At LHC  the values of signal
    strength
    $\mu_f \equiv \sigma(pp\longrightarrow H + X)*Br(H \longrightarrow
    f)/()_{SM}$ are measured.  What will the change in $\mu_f$ be in case of the
    fourth generation?

  \subsection{Why $N_q = N_l$?}

   $N_q = N_l$ in order to compensate chiral anomalies, which violate conservation
  of gauge axial currents,  making theory nonrenormalizable.

  The case of QED:
  \begin{center}
    \includegraphics[width=.5\textwidth]{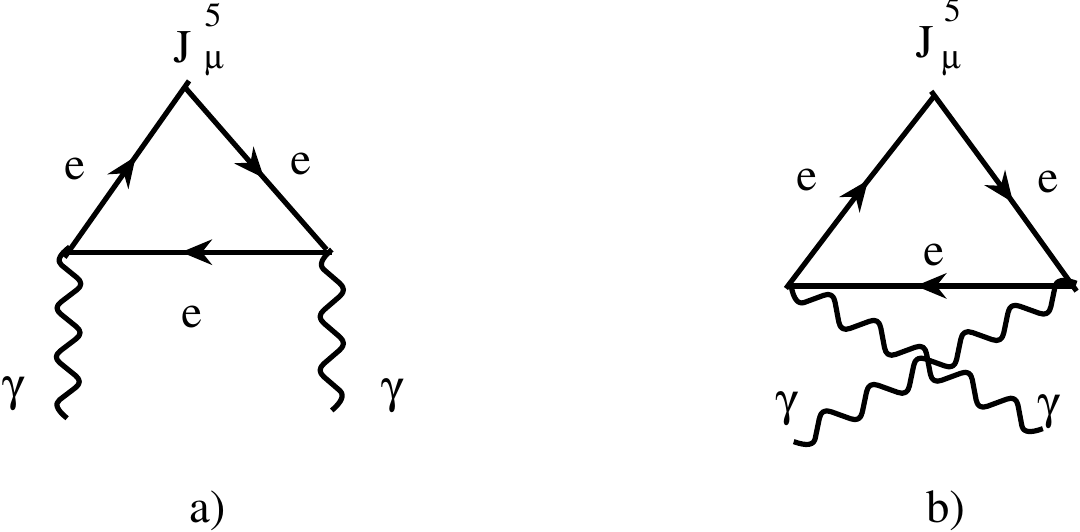}
  \end{center}

  Unlike QED, $SU(2)_L \times U(1)$ gauge invariant Standard Model (SM) \cite{0} deals with Weyl fermions. Thus the gauge bosons $A_i$ and
  $B$ interact with axial currents. In each generation the quarkonic and
  leptonic $A_i^2 B$ and $B^3$ triangles compensate each other, that
  is why $N_q$ should be equal to $N_l$.

  \bigskip
  Problem 2

    Prove that the quarkonic triangles cancel the leptonic ones when
    $Q_e=-Q_p$ (so hydrogen atoms are neutral) and
    $Q_n=Q_{\nu} = 0$ (thus neutrino and neutron are
    neutral).

 \section{ Cabibbo-Kobayashi-Maskawa (CKM) matrix, unitarity triangles}

 \subsection{The CKM matrix - where from?}

  In constructing the Standard Model Lagrangian the basic ingredients
  are:
  \begin{enumerate}
  \item gauge group,
  \item particle content,
  \item renormalizability of the theory.
  \end{enumerate}
  There is no such a building block in the Standard Model as CKM
  matrix in charged current quark interactions.

  This is the SM lagrangian:
  \begin{eqnarray}
 {\cal L}_{\rm SM} = -\frac{1}{2}{\rm tr} G_{\mu\nu}^2 - \frac{1}{2}{\rm tr} A_{\mu\nu}^2 -
  \frac{1}{4} B_{\mu\nu}^2 + |D_\mu H|^2 - \frac{\lambda^2}{2} [H^+ H - \eta^2/2]^2 + \nonumber    \\
 + \bar{Q}_L^i \hat D Q_L^i + \bar u_R^i \hat D u_R^i + \bar d_R^i \hat D d_R^i +
  \bar L_L^i \hat D L_L^i + \bar l_R^i \hat D l_R^i + \bar N_R^i \hat\partial  N_R^i +  \\
  + \left[f_{ik}^{(u)} \bar Q_L^i u_R^k  H + f_{ik}^{(d)} \bar Q_L^i d_R^k \tilde H +
    f_{ik}^{(\nu)} \bar L_L^i N_R^k  H + f_{ik}^{(l)} \bar L_L^i l_R^k \tilde H + M_{ik} N_R^i C^+ N_R^k + c.c. \right],  \nonumber
  \label{4}
  \end{eqnarray}

 \begin{equation}
  \hat D \equiv D_\mu \gamma_\mu \; , \;\; D_\mu = \partial_\mu - ig_s G_\mu^i \lambda_i/2
  - ig A_\mu^i \sigma_i/2 - ig^\prime B_\mu Y/2,
\label{5}
  \end{equation}

  where we suppose that neutrinos get masses by the see-saw mechanism.

   CKM matrix originates from Higgs field interactions with
  quarks.

  (All quark fields are primed: $Q_L \rightarrow Q_L^{'}, u_R \rightarrow u_R^{'},... $)

 \subsection{CKM matrix originates from Higgs field interactions with quarks.}

  The piece of the Lagrangian from which the up quarks get their
  masses looks like:

\begin{equation}
\Delta{\cal L}_{\rm up} = f_{ik}^{(u)} \bar
  Q_L^{i^\prime} u_R^{k^\prime} H + {\rm c.c.} \; , \;\; i,k = 1,2,3
  \;\; ,
\label{6}
  \end{equation}
  where

\begin{equation}
Q_L^{1^\prime} = \left(
    \begin{array}{cc} u^\prime
      \\ d^\prime \\ \end{array} \right)_L \; , \;\; Q_L^{2^\prime} =
  \left(
    \begin{array}{cc}
      c^\prime \\ s^\prime \\ \end{array} \right)_L \; , \;\;
  Q_L^{3^\prime} = \left(
    \begin{array}{cc} t^\prime \\ b^\prime \\ \end{array} \right)_L \;\; ;
\label{7}
\end{equation}
\begin{equation}
u_R^{1^\prime} = u_R^\prime \; , \;\; u_R^{2^\prime} =
  c_R^\prime \; , \;\; u_R^{3^\prime} = t_R^\prime
\label{8}
\end{equation}
  and $H$ is the higgs doublet:

\begin{equation}
H = \left(\begin{array}{cc} H^0
                 \\ H^-
               \end{array}\right).
\label{9}
\end{equation}

The piece of the Lagrangian which is responsible for the down
  quark masses looks the same way:
\begin{equation}
 \Delta{\cal L}_{\rm down} =
  f_{ik}^{(d)} \bar Q_L^{i^\prime} d_R^{k^\prime} \tilde H + {\rm
    c.c.} \; ,
\label{10}
\end{equation}
where
\begin{equation}
d_R^{1^\prime} = d_R^\prime \; ,
  \;\; d_R^{2^\prime} = s_R^\prime \; , \;\; d_R^{3^\prime} =
  b_R^\prime \;\; {\rm and} \;\; \tilde H_a = \varepsilon_{ab} H_b^*
  \; ,
  \label{11}
 \end{equation}
 \begin{equation}
 \varepsilon_{ab} = \left(
    \begin{array}{rl}
      0 & 1
      \\ -1 & 0
    \end{array} \right) \; .
  \label{12}
  \end{equation}
  After $SU(2) \times U(1)$ symmetry breaking by the Higgs field
  expectation value ${<H^0> = v}$, two mass matrices emerge:
  \begin{equation}
 M_{\rm up}^{ik} \bar u_L^{i^\prime} u_R^{k^\prime} + M_{\rm
    down}^{ik} \bar d_L^{i^\prime} d_R^{k^\prime} + c.c.
\label{13}
\end{equation}

The matrices $M_{\rm up}$ and $M_{\rm down}$ are
    arbitrary 3$\times$3 matrices; their matrix elements are complex
  numbers.  According to the very useful theorem, an arbitrary matrix
  can be written as a product of the hermitian and unitary
  matrices:
\begin{equation}
 M = UH \; , \;\; {\rm where} \;\; H = H^+ \;,
  \;\; {\rm and} \;\; UU^+ =1 \;\; ,
  \label{14}
  \end{equation}
  (do not mix the hermitian matrix $H$ with the Higgs field!)
  which is analogous to the following representation of an arbitrary
  complex number:
  \begin{equation}
 a = e^{i\phi} |a| \;\; .
\label{15}
 \end{equation}

 Matrix $M$ can be diagonalized by 2 different unitary matrices
  acting from left and right:
  \begin{equation}
 U_L M U_R^+ = M_{\rm diag} = \left(
    \begin{array}{ccc} m_u & & 0 \\ & m_c & \\ 0 & & m_t
    \end{array} \right) \;\; ,
 \label{16}
 \end{equation}
 where $m_i$ are the real numbers (if matrix $M$ is hermitian ($M =
  M^+$) then we will get $U_L = U_R$, the case of hamiltonian in QM). Having these formulas in mind,
  let us rewrite the up-quarks mass term:
  \begin{equation}
 \bar   u_L^{i^\prime} M_{ik} u_R^{k^\prime} + c.c. \equiv \bar u_L^\prime
  U_L^+ U_L M U_R^+ U_R u_R^\prime + c.c. = \bar u_L M_{\rm diag}
  u_R + c.c. = \bar u M_{\rm diag} u \;\; ,
  \label{17}
  \end{equation}
  where
  we introduce the fields $u_L$ and $u_R$ according to the following
  formulas:
  \begin{equation}
u_L = U_L u_L^\prime \; , \;\; u_R = U_R u_R^\prime
  \;\; .
  \label{18}
  \end{equation}

  Applying the same procedure to matrix $M_{\rm down}$ we observe that
  it becomes diagonal as well in the rotated basis:
\begin{equation}
d_L = D_L d_L^\prime \; , \;\; d_R = D_R d_R^\prime \;\;   .
\label{19}
 \end{equation}

  Thus we start from the primed quark fields and get
    that they should be rotated by 4 unitary matrices $U_L$, $U_R$,
    $D_L$ and $D_R$ in order to obtain unprimed fields with diagonal
    masses.

  Since kinetic energies and interactions with the vector fields
  $A_\mu^3$, $B_\mu$ and gluons are proportional to the unit matrix,
  these terms remain diagonal in a new unprimed basis. The only term
  in the SM Lagrangian where matrices $U$ and $D$ show up is
  charged current interactions with the emission of
  $W$-boson:
  \begin{equation}
   \Delta{\cal L} = g W_\mu^+ \bar u_L^\prime \gamma_\mu d_L^\prime
  = gW_\mu^+ \bar u_L \gamma_\mu U_L D_L^+ d_L \;\; ,
  \label{20}
  \end{equation}
  and   the unitary matrix $V\equiv U_L D_L^+$ is called
  Cabibbo-Kobayashi-Maskawa (CKM) quark mixing matrix.

\subsection{Parametrization of the CKM matrix: angles, phases, unitarity
    triangles}

  $n\times n$ unitary matrix has $n^2/2$ complex or $n^2$ real
  parameters. The orthogonal $n\times n$ matrix is specified by
  $n(n-1)/2$ angles (3 Euler angles in case of $O(3)$). That is why
  the parameters of the unitary matrix are divided between phases and
  angles according to the following relation:
  \begin{equation}
  \begin{array}{cccc} n^2 = & {\frac{n(n-1)}{2}} & + &
    \frac{n(n+1)}{2} \;\; . \\ &&& \\ & {\rm angles}&
    & {\rm phases}
  \end{array}
  \label{21}
  \end{equation}
  Are all these phases physical observables or, in other words, can
    they be measured experimentally?

  The answer is ``no'' since we can perform phase rotations of quark
  fields ($u_L \to e^{i\zeta}u_L$, $d_L\to e^{i\xi} d_L$ ...)
  removing in this way $2n-1$ phases of the CKM
  matrix. The number of unphysical phases equals the number of up and
  down quark fields minus one. The simultaneous rotation of all
  up-quarks on one and the same phase multiplies
  all the matrix elements of matrix $V$ by (minus) this phase. The rotation of all
  down-quark fields on one and the same phase acts on $V$ in the same
  way. That is why the number of the ``unremovable'' phases of matrix
  $V$ is decreased by the number of possible rotations of up and down
 quarks
  minus one.

  \bigskip
  Finally for the number of observable phases we get:
  \begin{equation}
 \frac{n(n+1)}{2} - (2n-1) = \frac{(n-1)(n-2)}{2} \;\; .
 \label{22}
 \end{equation}

  As you see, for the first time one observable phase arrives in the
  case of 3 quark-lepton generations.

\subsection{A bit of history}

  Introduced in 1963 by Cabibbo angle $\theta_c$ \cite{1} in a modern language
  mixes $d$- and $s$-quarks in the expression for the charged quark
  current:
  \begin{equation}
 J_\mu^+ = \bar u \gamma_\mu(1+\gamma_5)[d \cos\theta_c +
  s\sin\theta_c] \;\; .
  \label{23}
\end{equation}
  In this way he related the suppression of the strange particles weak
  decays to the smallness of angle $\theta_c$,
  $\sin^2\theta_c \approx 0.05$.\footnote{Earlier in the framework of
  "eightfold way`` such a suppression of the charged strange current was
  discussed by Gell-Mann \cite{2}.}  In order to explain the suppression
  of $K^0-\bar K^0$ transition the GIM mechanism (and c-quark) was
  suggested in 1970 \cite{3}.  After the discovery of a $J/\Psi$-meson made from
  ($c \bar c$) quarks in 1974 it
  was confirmed that 2 quark-lepton generations exist. The mixing of
  two quark generations is described by the unitary 2$\times$2 matrix
  parametrised by one angle and zero observable phases. This angle is
  Cabibbo angle.

  However, even before the $c$-quark discovery in 1973 Kobayashi and
  Maskawa noticed that one of the several ways to implement
  CP-violation in the Standard Model is to postulate the existence of
  3 quark-lepton generations since for the first time the observable
  phase shows up for $n=3$ \cite{4}. At that time CPV was known only in neutral
  $K$-meson decays and to test KM mechanism one needed other
  systems. Almost 30 years after KM model had been suggested it was
  confirmed in $B$-meson decays.

  Here is the CKM matrix
\begin{equation}
\overline{(u c t)_L} \left(
    \begin{array}{lll} V_{ud} & V_{us} &
                                         V_{ub}
      \\ V_{cd} & V_{cs} & V_{cb} \\ V_{td} & V_{ts} & V_{tb}
    \end{array} \right)
  \left( \begin{array}{c} d \\ s \\ b \end{array} \right)_L ,\;\;
  \label{24}
  \end{equation}

  and it's standard parametrization looks like:
  \begin{equation}
  V = R_{23} \times R_{13} \times R_{12} \;\; ,
  \label{25}
  \end{equation}
  \begin{equation}
  R_{23} = \left(
    \begin{array}{ccc}
      1 & 0 & 0 \\
      0 & c_{23} & s_{23} \\
      0 & -s_{23} & c_{23}
    \end{array}
  \right),  \;
    R_{13} =
  \left(
    \begin{array}{ccc}
      c_{13} & 0 & s_{13} e^{-i\delta} \\
      0 & 1 & 0 \\
      -s_{13}e^{i\delta} & 0 & c_{13}
    \end{array}
  \right),
  \;
  R_{12} = \left(
    \begin{array}{ccc}
      c_{12} & s_{12} & 0 \\
      -s_{12} & c_{12} & 0 \\
      0 & 0 & 1
    \end{array}
  \right),
  \label{27}
  \end{equation}
    and, finally:
 \begin{equation}
 V = \left(
    \begin{array}{ccc}
      c_{13} c_{12} & c_{13} s_{12} & s_{13} e^{-i\delta} \\
      -c_{23} s_{12} -s_{23}s_{13} c_{12} e^{i\delta} & c_{23} c_{12}
                                                        -s_{12} s_{13}
                                                        s_{23}e^{i\delta}
                                    & s_{23} c_{13} \\
      s_{12} s_{23} -c_{12} c_{23} s_{13}e^{i\delta} & -s_{23} c_{12}
                                                       -c_{23} s_{13}
                                                       s_{12}
                                                       e^{i\delta} & c_{23} c_{13}
    \end{array}
  \right) \; .
  \label{28}
\end{equation}

\subsection{Wolfenstein parametrization}
  Let us introduce new
  parameters $\lambda$, $A$, $\rho$ and $\eta$ according to the
  following definitions:
 \begin{equation}
 \lambda \equiv s_{12} , \;\;  A\equiv
  \frac{s_{23}}{s_{12}^2}  , \;\; \rho =
  \frac{s_{13}}{s_{12}s_{23}}\cos\delta , \;\; \eta = \frac{s_{13}}{s_{12}s_{23}}\sin\delta \;\; ,
  \label{29}
  \end{equation}
  and get the expressions for $V_{ik}$ through $\lambda$, $A$, $\rho$
  and $\eta$:
  \begin{equation}
  V = \left(
    \begin{array}{lll}
      V_{ud} & V_{us} & V_{ub}\\
      V_{cd} & V_{cs} & V_{cb}\\
      V_{td} & V_{ts} & V_{tb}
    \end{array} \right)
  \approx \left(
    \begin{array}{ccc}
      1-\lambda^2/2 & \lambda & A\lambda^3(\rho -i\eta) \\
      -\lambda -iA^2 \lambda^5 \eta & 1-\lambda^2/2 & A\lambda^2 \\
      A\lambda^3(1-\rho -i\eta) & -A\lambda^2-iA\lambda^4 \eta & 1 \end{array} \right).
      \label{31}
      \end{equation}

  In the last expression the expansion in powers of $\lambda$ is made.

  The last form of CKM matrix is very convenient for qualitative
  estimates \cite{5}. Approximately we have: $ \lambda \approx
    0.225, A \approx 0.83, \eta \approx 0.36, \rho \approx 0.15$.

\subsection{Unitarity triangles; FCNC}

  The unitarity of the matrix $V$ ($V^+V=1$) leads to the following six
  equations that can be drawn as triangles on a complex plane (under
  each term in these equations the power of $\lambda$ entering it, is
  shown):
\begin{equation}
  \begin{array}{cccccc}
    V_{ud}^*V_{us} & + & V_{cd}^* V_{cs} & + & V_{td}^* V_{ts} & = 0 \;\;\;\;\;\;\;\;\;\;s\rightarrow d \\
    \sim\lambda & & \sim\lambda & & \sim\lambda^5 &
  \end{array}
\label{32}
\end{equation}
\begin{equation}
 \begin{array}{cccccc}
    V_{ud}^* V_{ub} & + & V_{cd}^* V_{cb} & + & V_{td}^* V_{tb} & = 0 \;\; \;\;\;\;\;\;\;\; b\rightarrow d\\
    \sim\lambda^3 & & \sim\lambda^3 & & \sim\lambda^3
  \end{array}
\label{33}
\end{equation}
\begin{equation}
 \begin{array}{cccccc}
    V_{us}^* V_{ub} & + & V_{cs}^* V_{cb} & + & V_{ts}^* V_{tb} & = 0 \;\; \;\; \;\;\;\;\;\;\;\; b\rightarrow s\\
    \sim\lambda^4 & & \sim\lambda^2 & & \sim\lambda^2 &
  \end{array}
  \label{34}
  \end{equation}
\begin{equation}
  \begin{array}{cccccc}
    V_{ud} V_{cd}^* & + & V_{us} V_{cs}^* & + & V_{ub} V_{cb}^* & = 0 \;\; \;\; \;\;\;\;\;\;\;\; c\rightarrow u \\
    \sim\lambda & & \sim\lambda & & \sim\lambda^5 &
  \end{array}
\label{35}
\end{equation}
\begin{equation}
  \begin{array}{cccccc}
    V_{ud} V_{td}^* & + & V_{us} V_{ts}^* & + & V_{ub} V_{tb}^* & = 0 \;\;  \\
    \sim\lambda^3 & & \sim\lambda^3 & & \sim\lambda^3 &
  \end{array}
\label{36}
\end{equation}
\begin{equation}
  \begin{array}{cccccc}
    V_{cd} V_{td}^* & + & V_{cs} V_{ts}^* & + & V_{cb} V_{tb}^* & = 0\;\;  \\
    \sim\lambda^4 & & \sim\lambda^2 & & \sim\lambda^2 &
  \end{array}
  \label{37}
  \end{equation}

  Among these triangles four are almost degenerate: one side is much
  shorter than two others, and two triangles have all three sides of
  more or less equal lengths, of the order of $\lambda^3$. These two
  nondegenerate triangles almost coincide.

\begin{center}

\begin{figure}[h]

\hspace{45mm} \includegraphics[width=.5\textwidth]{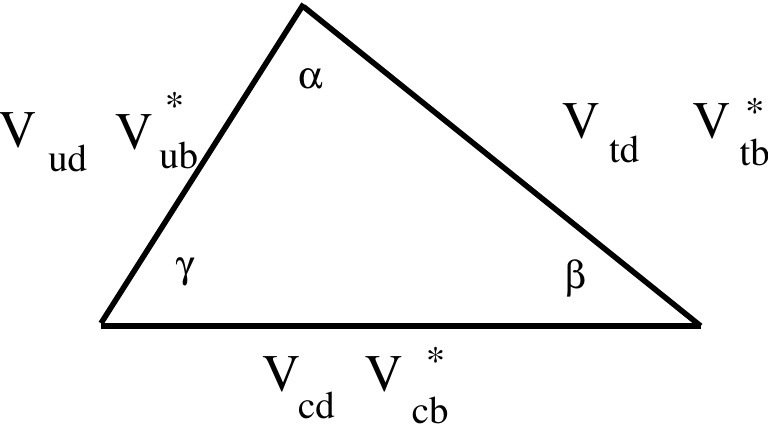}

      \caption{Unitarity triangle}
      \label{fig:ut}

    \end{figure}

  \end{center}

  So, as a result we have only one nondegenerate unitarity triangle;
  it is usually defined by a complex conjugate of our equation:
  \begin{equation}
  V_{ud}V_{ub}^* +V_{cd} V_{cb}^* +V_{td}V_{tb}^* = 0 \;\;
  \label{38}
  \end{equation}
  and it is shown in \Figure\ref{fig:ut}. It has the angles which
  are called $\beta$, $\alpha$ and $\gamma$.
  They are determined from CPV asymmetries in
  $B$-mesons decays.

  Looking at \Figure\ref{fig:ut} one can easily obtain the following formulas:
  \begin{equation}
  \beta = \pi - \arg\frac{V_{tb}^* V_{td}}{V_{cb}^* V_{cd}} =
  \phi_1
  \label{39}
  \end{equation}
  \begin{equation}
  \alpha = \arg\frac{V_{tb}^* V_{td}}{-V_{ub}^* V_{ud}} = \phi_2
  \label{40}
  \end{equation}
  \begin{equation}
  \gamma = \arg\frac{V_{ub}^* V_{ud}}{-V_{cb}^* V_{cd}} = \phi_3
  \label{41}
  \end{equation}

    \begin{itemize}
    \item Angle $\beta$ was measured through time dependent CPV
      asymmetry in $B_d \to {\rm charmonium} \; K^0$ decays,
    \item Angle $\alpha$ was measured from CPV asymmetries in
      $B_d \to\pi \pi, \rho \rho$ and $\pi \rho$ decays,
    \item $B^{\pm}$ decays are used to determine angle $\gamma$.
    \end{itemize}

  Multiplying any quark field by an arbitrary phase and absorbing it
  by CKM matrix elements we do not change some unitarity triangles,
  while the others are rotating as a whole, preserving their shapes
  and areas.  For the area of any of unitarity triangle we get:
  \begin{equation}
  A=1/2 {\rm Im}(a\cdot b^*) = 1/2|a|\cdot|b|\cdot\sin\alpha \;\; ,
  \label{42}
  \end{equation}
  where $a$ and $b$ are the sides of the triangle.

  \bigskip

  Problem 3

    Prove that the areas of all unitarity triangles are the same.
    \emph{Hint:} Use equations which define unitarity triangles.
  \subsection{Cecilia Jarlskog's invariant}
  The area of unitarity triangles contains an important information
  about the properties of CKM matrix.

  CPV in the SM is proportional to this area, which equals $1/2$ of
  the Jarlskog invariant $J$ \cite{6}.

    Writing $J= Im(V_{ud}V_{ub}^*V_{cd}^* V_{cb})$ we see, that $J$ is
  not changed when quark fields are multiplied by arbitrary phases.

  \bigskip

  The source of CPV in the SM is the phase $\delta$ - this is a correct
  statement; BUT it
  is like a phantom. If somebody says that the source of CPV is the
  phase of $V_{td}$, then another one can rotate $d$-quark, or
  $t$-quark, or both making $V_{td}$ real.

  However, there is an invariant quantity, which is not
    a phantom - $J$.

\section{CP, CP violation}

\subsection{CP: history}

 Landau thought that space-time symmetries of a Lagrangian should be
  that of an empty space. Indeed, from a shift symmetry we deduce energy and momentum
  conservation, from rotation symmetry - angular momentum conservation.  In
  1956 Lee and Yang (in order to solve $\theta-\tau$ problem) suggested
  that P-parity is broken in weak interactions \cite{7}.

  This was unacceptable for Landau: empty space has left-right
  interchange symmetry, so a Lagrangian should have it as well. Then
  Ioffe, Okun and Rudik noted that Lee and Yang's theory violates
  charge conjugation symmetry (C) as well, while CP is conserved
  explaining the difference of life times of $K_L-$ and $K_S-$ mesons
  \cite{8} a-la Gell-Mann and Pais \cite{9} but with CP replacing $C$. $C$-parity violation in weak interactions was discussed in \cite{888} as well.

  Just at this point Landau found the way to resurrect P-invariance
  stating that the theory should be invariant under the product of P
  reflection and C conjugation. He called this product the combined
  inversion and according to him it should substitute $P$-inversion
  broken in weak interactions. In this way the theory should be
  invariant when together with changing the sign of the coordinates,
  $\bar r \to -\bar r$, one changes an electron to positron, proton to
  antiproton and so on. Combined parity instead of parity.

  It is clearly seen from 1957 Landau paper that CP-invariance should
  become a basic symmetry for physics in general and weak interactions
  in particular \cite{10}.

  Nevertheless  L.B.Okun considered the search for
    $K_L \rightarrow 2\pi$ decay to be one of the most important
    problems in weak interactions \cite{11}.

  The notion of CP appears to be so important, that more than 60 years
   later you are listening
   to the lectures on CPV.

\subsection{PV}

  Landau's answer to the question ``Why is parity violated in weak
  interactions'' was: because CP, not P is the fundamental symmetry of
  nature.

  A modern answer to the same question is: because in
  P-invariant theory with the Dirac fermions the gauge invariant mass
  terms can be written for quarks and leptons which are not protected
  from being of the order of $M_{\rm GUT}$ or $M_{\rm Planck}$. So in
  order to have our world made from light particles P-parity should be
  violated, thus Weyl fermions should be used.

   \subsection{CPV}

  $K_L \to 2\pi$ decay discovered in 1964 by Christenson, Cronin,
  Fitch and Turlay \cite{12} occurs due to CPV in the mixing of neutral kaons
  ($\tilde\varepsilon \neq 0$). Only thirty years later the second
  major step was done: direct CPV was observed in kaon
  decays \cite{13}:
  \begin{equation}
  \frac{\Gamma(K_L \to \pi^+ \pi^-)}{\Gamma(K_S \to \pi^+ \pi^-)}
  \neq \frac{\Gamma(K_L \to \pi^0 \pi^0)}{\Gamma(K_S \to \pi^0 \pi^0)}
  \; , \;\; \varepsilon^\prime \neq 0   \;\; .
  \label{43}
  \end{equation}

  \bigskip

  In the year 2001 CPV was for the first time observed beyond the
  decays of neutral kaons: the time dependent CP-violating asymmetry
  in $B^0$ decays was measured \cite{14}:
  \begin{equation}
  a(t)=\frac{dN(B^0 \to J/\Psi K_{S(L)})/dt-dN(\bar B^0 \to J/\Psi
    K_{S(L)})/dt}{dN(B^0 \to J/\Psi K_{S(L)})/dt+ dN(\bar B^0 \to J/\Psi
    K_{S(L)})/dt}\neq 0 \;\; .
    \label{44}
    \end{equation}

  \bigskip

  Finally, in 2019 direct CPV was found in $D^0(\bar D^0)$
  decays to $\pi^+\pi^-(K^+K^-)$ \cite{15}.

  Since 1964 we have known that there is no symmetry between particles
  and antiparticles. In particular, the $C$-conjugated partial widths
  are different:
  \begin{equation}
\Gamma(A \to BC) \neq \Gamma(\bar A \to \bar B \bar C) \;\; .
\label{45}
\end{equation}
However, CPT (deduced from the invariance of the theory under
  4-dimen\-sional rotations) remains intact. That is why the total
  widths as well as the masses of particles and antiparticles are
  equal:
  \begin{equation}
M_A = M_{\bar A} \;, \;\; \Gamma_A = \Gamma_{\bar A} ~~~~
        ({\rm CPT}) \;\; .
\label{46}
\end{equation}

The consequences of CPV can be divided into macroscopic and
  microscopic. CPV is one of the three famous Sakharov's conditions to
  get a charge
  nonsymmetric Universe as a result of evolution of a charge symmetric one \cite{16}. In these lectures we will not discuss this very
  interesting branch of physics, but will deal with CPV in particle
  physics where the data obtained up to now confirm Kobayashi-Maskawa
  model of CPV. New data which should become available in coming years
  may as well disprove it clearly demonstrating the necessity of
  physics beyond the Standard Model.

\subsection{CPV and complex couplings }

The next question I would like to discuss is why the phases are
relevant for CPV. In the SM  charged currents are left-handed:
\begin{equation}
\Delta{\cal L} = g
\bar u_L \gamma_\mu V d_L W_\mu +  g\bar d_L \gamma_\mu V^+ u_L W_\mu^* \;\; .
\label{47}
\end{equation}
Under space inversion (P)
they become right-handed. Under charge conjugation (C) left-handed
charged currents become right-handed as well and field operators become
complex conjugate.

So, weak interactions are P- and C-odd.

However,  CP transforms the left-handed current to left-handed,
so the theory can be CP-even. If all coupling constants in
the SM Lagrangian were real then, being hermitian, Lagrangian would be CP invariant.

Since coupling constants of charged currents are complex (there is the CKM matrix $V$)
CP invariance is violated.
But when complex phases can be absorbed by field operators redefinition there is no CPV (the cases
of one or two quark-lepton generations).
\begin{equation}
{\cal L}_W = \frac{g}{\sqrt 2} \bar u \gamma_\mu \frac{1+\gamma_5}{2} V d W_\mu +
\frac{g}{\sqrt 2} \bar d \gamma_\mu \frac{1+\gamma_5}{2} V^+ u W_\mu^*
\label{48}
\end{equation}
\begin{equation}
P\psi = i\gamma_0\psi \; , \;\; P(W_0, W_i) = (W_0, -W_i)
\label{49}
\end{equation}
\begin{equation}
\bar u(\gamma_0, \gamma_i)d \to \bar u(\gamma_0, -\gamma_i)d
\label{50}
\end{equation}
\begin{equation}
\bar u(\gamma_0 \gamma_5, \gamma_i \gamma_5) d \to
\bar u(-\gamma_0 \gamma_5, \gamma_i\gamma_5)d
\label{51}
\end{equation}
\begin{equation}
{\cal L}_W^P = \frac{g}{\sqrt 2} \bar u \gamma_\mu \frac{1-\gamma_5}{2}VdW_\mu +
\frac{g}{\sqrt 2} \bar d \gamma_\mu \frac{1-\gamma_5}{2} V^+ u W_\mu^* \;\; ,
\label{52}
\end{equation}
\begin{equation}
C\psi = \gamma_2 \gamma_0 \bar\psi \; , \;\; C(W_0, W_i) = -(W_0^*, W_i^*)
\label{53}
\end{equation}
\begin{equation}
{\cal L}_W^C = \frac{g}{\sqrt 2} \bar d \gamma_\mu \frac{1-\gamma_5}{2}V^T u
W_\mu^* + \frac{g}{\sqrt 2} \bar u \gamma_\mu \frac{1-\gamma_5}{2} V^* d W_\mu
\label{54}
\end{equation}
\begin{equation}
{\cal L}_W^{\rm CP} = \frac{g}{\sqrt 2} \bar d \gamma_\mu \frac{1+\gamma_5}{2}
V^Tu W_\mu^* + \frac{g}{\sqrt 2} \bar u \gamma_\mu \frac{1+\gamma_5}{2}V^* d W_\mu
\label{55}
\end{equation}

Real $V$: ${\cal L}_W^{\rm CP} = {\cal L}_W$, no CPV.

Complex $V$: it cannot be made real by fields redefinition $u_i\to e^{i\alpha_i} u_i$, $d_j \to e^{i\beta_j}d_j$
when
$N_{\rm gen} \geq 3$ -- all phases cannot be eliminated and CP is violated.

\bigskip

\section{$\mbox{\boldmath$M^0 - \bar M^0$}$ mixing; CPV in
mixing}

In order to mix, a meson must be neutral and not
coincide with its antiparticle.
There are four such pairs:
\begin{eqnarray}
K^0(\bar s d) - \bar
K^0(s\bar d) \; , \;\; D^0(c\bar u) - \bar D^0(\bar c u) \;\; , \nonumber \\
B_d^0(\bar b d) - \bar B_d^0(b\bar d) \;\;  {\rm and} \;\;
B_s^0(\bar b s) - \bar B_s^0(b \bar s) \;\; .
\label{56}
\end{eqnarray}

\begin{figure}[!htb]
  \centering
  \begin{tikzpicture}[scale=1.3]
    \tikzset{every node/.append style={font=\normalsize}}
      \coordinate (A) at (0,0);
      \coordinate (B) at (1,0);
      \coordinate (C) at (3,0);
      \coordinate (D) at (4,0);
      \coordinate (E) at (0,2);
      \coordinate (F) at (1,2);
      \coordinate (G) at (3,2);
      \coordinate (H) at (4,2);
      \coordinate (S1) at (0,1);
      \coordinate (S2) at (4,1);
      \draw[electron] (A) to node[below]{$s$} (B);
      \draw[electron] (B) to node[below]{$u,c,t$} (C);
      \draw[electron] (C) to node[below]{$d$} (D);
      \draw[electron] (F) to node[above]{$ d$} (E);
      \draw[electron] (G) to node[above]{$ u,~ c,~ t$} (F);
      \draw[electron] (H) to node[above]{$ s$} (G);
      \draw[photon] (B) to node[left]{$W$} (F);
      \draw[photon] (C) to node[right]{$W$} (G);
      \draw (S1) node[left]{$K^{0}$};
      \draw (S2) node[right]{$\bar K^{0}$};
      \filldraw (B) circle (1.5pt);
      \filldraw (C) circle (1.5pt);
      \filldraw (F) circle (1.5pt);
      \filldraw (G) circle (1.5pt);
     \end{tikzpicture}
  \caption{$K^0 - \bar K^0$ transition.}
  \label{fig:kmix}
\end{figure}
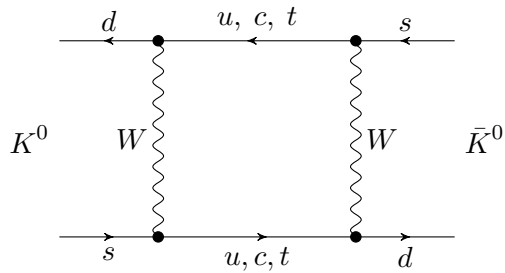

Mixing occurs in the second order in weak interactions through the
box diagram which is shown in \Figure\ref{fig:kmix} for $K^0 - \bar K^0$ pair.

The effective $2\times 2$ Hamiltonian $H$ is used to describe the
meson-antimeson mixing. It is most easily written in the following
basis:
\vspace{-0.5em}
\begin{equation}
M^0 = \left(\begin{array}{c} 1 \\ 0
\end{array} \right),~ \bar M^0 = \left(\begin{array}{c} 0 \\ 1
\label{57}
\end{array} \right).
\end{equation}

The meson-antimeson system evolves according
to the Shroedinger equation with this effective Hamiltonian which
is not hermitian since it takes meson decays into account. So, $H=
M - \frac{i}{2} \Gamma$,  where both $M$ and $\Gamma$ are hermitian.

According to CPT invariance the diagonal elements of $H$ are
equal:
\begin{equation}
\langle M^0 \mid H\mid M^0 > = <\bar M^0 \mid H \mid \bar M^0
\rangle \;\; .
\label{58}
\end{equation}

Substituting into the Shroedinger equation
\begin{equation}
i\frac{\partial\psi}{\partial t} = H\psi
\label{59}
\end{equation}
$\psi$ -- function in the following form:
\begin{equation}
\psi = \left(\begin{array}{c} p
\\ q \end{array}\right) e^{-i\lambda t}
\label{60}
\end{equation}
we come to the following equation:
\begin{equation}
\left( \begin{array}{cc}M - \frac{i}{2} \Gamma & M_{12}
-\frac{i}{2} \Gamma_{12} \\ ~ &  ~\\ M_{12}^* - \frac{i}{2}
\Gamma_{12}^* & M -\frac{i}{2} \Gamma  \end{array} \right) \left(
\begin{array}{c} p \\  ~  \\ q \end{array} \right) = \lambda \left(
\begin{array}{c} p \\ ~  \\ q \end{array} \right)
\label{61}
\end{equation}
from which for eigenvalues ($\lambda_\pm$) and eigenvectors
($M_\pm$) we obtain:
\begin{equation}
\lambda_{\pm} = M -\frac{i}{2}\Gamma \pm
\sqrt{(M_{12} - \frac{i}{2}\Gamma_{12})(M_{12}^* -
\frac{i}{2}\Gamma_{12}^*)} \; ,
\label{62}
\end{equation}
\begin{equation}
\left\{
\begin{array}{l} M_+ = pM^0 + q \bar M^0 \\ M_- = pM^0 - q \bar
M^0 \end{array} \right. \;\; , \;\; \frac{q}{p} =
\sqrt{\frac{M_{12}^* - \frac{i}{2}\Gamma_{12}^*}{M_{12}
-\frac{i}{2}\Gamma_{12}}} \;\; .
\label{63}
\end{equation}

If there is no CPV in mixing, then:
\begin{eqnarray}
\langle M^0\mid H \mid \bar M^0 \rangle
= \langle \bar M^0 \mid H \mid M^0 \rangle  \;\; , \nonumber \\
M_{12} -
\frac{i}{2} \Gamma_{12} = M_{12}^* - \frac{i}{2} \Gamma_{12}^*
\;\; ,
\label{64}
\end{eqnarray}
and
\begin{equation}
\frac{q}{p} = 1
\;\; , \;\; <M_+ \mid M_-
> = 0 \;\; ({\rm in} \;{\rm case}\; {\rm of}\; {\rm kaons} \; M_+=K^0_1,\; M_-=K^0_2).
\label{65}
\end{equation}

However, even if the phases of $M_{12}$ and $\Gamma_{12}$ are nonzero but equal (modulo $\pi$) we
can eliminate this common phase rotating $M^0$.

We observe the one-to-one correspondence between CPV in mixing and
nonorthogonality of the eigenstates $M_+$ and $M_-$. According to
Quantum Mechanics if two hermitian matrices $M$ and $\Gamma$
commute, then they have a common orthonormal basis. Let us calculate
the commutator of $M$ and $\Gamma$:
\begin{equation}
[M, \Gamma] =\left(
\begin{array}{cc} M_{12} \Gamma_{12}^* - M_{12}^* \Gamma_{12} & 0
\\ ~ \\ 0 & M_{12}^* \Gamma_{12} - M_{12} \Gamma_{12}^*
\end{array}\right) \; .
\label{66}
\end{equation}
It equals zero if the phases of $M_{12}$ and $\Gamma_{12}$
coincide (modulo $\pi$). So, for $[M \Gamma] =0$ we get $\mid
q/p\mid =1$, $<M_+ \mid M_- > =0$ and there is no CPV in the
meson-antimeson mixing. And vice versa.

\bigskip

Problem 4

CPV in kaon mixing. According to the box diagram which describes $K^0 -\bar K^0$ mixing  $\Gamma_{12}\sim (V_{ud}^*V_{us})^2$. Find an analogous expression for $M_{12}$. Use unitarity of the matrix $V$ and eliminate $V_{cd}^*V_{cs}$ from $M_{12}$. Observe that the quantity $ M_{12} \Gamma_{12}^* - M_{12}^* \Gamma_{12}$ is proportional to
the Jarlskog invariant $J= Im(V_{ud}^*V_{us}V_{td} V_{ts}^*)$.

\bigskip

Introducing quantity $\tilde\varepsilon$ according to the
following definition:
\begin{equation}
\frac{q}{p} =
\frac{1-\tilde\varepsilon}{1+\tilde\varepsilon} \;\; ,
\label{67}
\end{equation}
we see that if $Re~\tilde\varepsilon \neq 0$, then CP is
violated. For the eigenstates we obtain:
\begin{eqnarray}
M_+ =
\frac{1}{\sqrt{1+ \mid \tilde\varepsilon \mid^2}} \left[\frac{M^0
+\bar M^0}{\sqrt{2}} + \tilde\varepsilon \frac{M^0 - \bar
M^0}{\sqrt{2}} \right] \;\; , \nonumber \\
M_- = \frac{1}{\sqrt{1+ \mid
\tilde\varepsilon \mid^2}} \left[\frac{M^0 -\bar M^0}{\sqrt{2}} +
\tilde\varepsilon \frac{M^0 +\bar M^0}{\sqrt{2}} \right] \;\; .
\label{68}
\end{eqnarray}

If CP is conserved, then $Re~\tilde\varepsilon = 0$, $M_+$ is CP
even and $M_-$ is CP odd. If CP is violated in mixing, then
$Re~\tilde\varepsilon \neq 0$ and $M_+$ and $M_-$ get admixtures
of the opposite CP parities and become nonorthogonal.

\bigskip

\section{Neutral kaons: mixing ($\Delta m_{LS}$) and CPV in mixing
    ($\tilde\varepsilon$)}

\subsection{\boldmath$K^0 - \bar K^0$ mixing, \boldmath $\Delta
m_{LS} $}

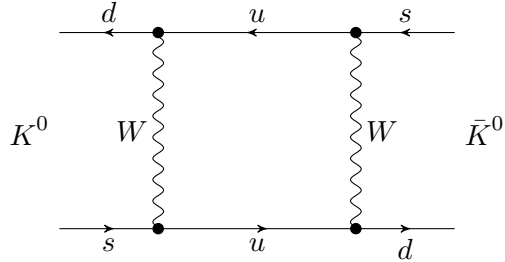
\begin{figure}[h]
  \centering
  \begin{tikzpicture}[scale=1.3]
    \tikzset{every node/.append style={font=\normalsize}}
      \coordinate (A) at (0,0);
      \coordinate (B) at (1,0);
      \coordinate (C) at (3,0);
      \coordinate (D) at (4,0);
      \coordinate (E) at (0,2);
      \coordinate (F) at (1,2);
      \coordinate (G) at (3,2);
      \coordinate (H) at (4,2);
      \coordinate (S1) at (0,1);
      \coordinate (S2) at (4,1);
      \coordinate (C1) at (2,0);
      \coordinate (C2) at (2,2);
      \draw[electron] (A) to node[below]{$s$} (B);
      \draw[electron] (B) to node[below]{$u$} (C);
      \draw[electron] (C) to node[below]{$d$} (D);
      \draw[electron] (F) to node[above]{$d$} (E);
      \draw[electron] (G) to node[above]{$u$} (F);
      \draw[electron] (H) to node[above]{$s$} (G);
      \draw[photon] (B) to node[left]{$W$} (F);
      \draw[photon] (C) to node[right]{$W$} (G);
      \draw (S1) node[left]{$K^{0}$};
      \draw (S2) node[right]{$\bar K^{0}$};
      \filldraw (B) circle (1.5pt);
      \filldraw (C) circle (1.5pt);
      \filldraw (F) circle (1.5pt);
      \filldraw (G) circle (1.5pt);
     \end{tikzpicture}
  \caption{The diagram which contributes to $\Gamma_{12}$.}

 \label{fig:box1}
\end{figure}

$\Gamma_{12}$ for the $K^0 - \bar K^0$ system is given by the
absorptive part of the diagram in \Figure\ref{fig:box1}. With our choice of CKM matrix $V_{us}$
and $V_{ud}$ are real, so  $\Gamma_{12}$ is real.

\begin{figure}
  \centering
  \begin{tikzpicture}[scale=1.3]
    \tikzset{every node/.append style={font=\normalsize}}
      \coordinate (A) at (0,0);
      \coordinate (B) at (1,0);
      \coordinate (C) at (3,0);
      \coordinate (D) at (4,0);
      \coordinate (E) at (0,2);
      \coordinate (F) at (1,2);
      \coordinate (G) at (3,2);
      \coordinate (H) at (4,2);
      \coordinate (S1) at (0,1);
      \coordinate (S2) at (4,1);
      \draw[electron] (A) to node[below]{$s$} (B);
      \draw[electron] (B) to node[below]{$u,c,t$} (C);
      \draw[electron] (C) to node[below]{$d$} (D);
      \draw[electron] (F) to node[above]{$ d$} (E);
      \draw[electron] (G) to node[above]{$ u,~ c,~ t$} (F);
      \draw[electron] (H) to node[above]{$ s$} (G);
      \draw[photon] (B) to node[left]{$W$} (F);
      \draw[photon] (C) to node[right]{$W$} (G);
      \draw (S1) node[left]{$K^{0}$};
      \draw (S2) node[right]{$\bar K^{0}$};
      \filldraw (B) circle (1.5pt);
      \filldraw (C) circle (1.5pt);
      \filldraw (F) circle (1.5pt);
      \filldraw (G) circle (1.5pt);
     \end{tikzpicture}
  \caption{The diagram which contributes to $M_{12}$.}
  \label{fig:box2}
\end{figure}
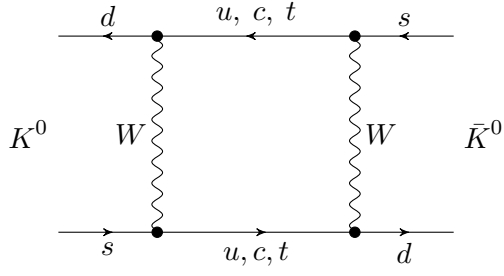

$M_{12}$ is given by a dispersive part of the diagram in \Figure\ref{fig:box2}.
Now all three up quarks should be taken into account.

To calculate this diagram it is convenient to implement GIM (Glashow-Illiopulos-Maiani)
compensation mechanism \cite{3} from the very beginning, subtracting zero
from the sum of the fermion propagators:
\begin{equation}
\frac{V_{us}
V_{ud}^*}{\hat p - m_u} + \frac{V_{cs} V_{cd}^*}{\hat p - m_c} +
\frac{V_{ts} V_{td}^*}{\hat p - m_t} - \frac{\sum\limits_i V_{is}
V_{id}^*}{\hat p} \;\; .
\label{69}
\end{equation}

Since $u$-quark is massless with good accuracy, $m_u \approx 0$,
then its propagator drops out and we are left with the modified $c$-
and $t$-quark propagators:
\begin{equation}
\frac{1}{\hat p - m_{c,t}}
\longrightarrow \frac{m_{c,t}^2}{(p^2 - m_{c,t}^2) \hat p} \;\; .
\label{70}
\end{equation}

The modified fermion propagators decrease in ultraviolet so
rapidly that one can calculate the box diagrams in the unitary
gauge, where $W$-boson propagator is $(g_{\mu\nu}-k_{\mu}k_{\nu}/M^2_W)
/(k^2-M_W^2)$

We easily get the following estimates for three remaining diagram
contributions to $M_{12}$:
\begin{eqnarray}
 (cc): & {\lambda^2(1-2i\eta
A^2 \lambda^4) G_F^2 m_c^2} \;\; , \nonumber \\
(ct): &
\lambda^6(1-\rho +i\eta) G_F^2 m_c^2
\ln(\frac{m_t}{m_c})^2 \;\; , \\
(tt): &
\lambda^{10}(1-\rho +i\eta)^2
 G_F^2 m_t^2 \;\; . \nonumber
 \label{71}
 \end{eqnarray}

Since $m_c \approx 1.3$ GeV and $m_t \approx 175$ GeV we observe
that the $cc$ diagram dominates in $Re M_{12}$ while $Im M_{12}$
is dominated by ($tt$) diagram.

$M_{12}$ is mostly real:
\begin{equation}
\frac{Im M_{12}}{Re M_{12}} \sim \lambda^8
\left(\frac{m_t}{m_c}\right)^2\sim 0.1 \;\; .
\label{72}
\end{equation}

The explicit calculation of the $cc$ exchange diagram gives:
\begin{equation}
{\cal L}_{\Delta s =2}^{\rm eff} = -\frac{g^4}{2^9 \pi^2
M_W^4}(\bar s \gamma_\alpha(1+\gamma_s)d)^2 \eta_1 m_c^2 V_{cs}^2
V_{cd}^ {*^2}\;\; ,
\label{73}
\end{equation}

where $g$ is SU(2) gauge
coupling constant, $g^2/8 M_W^2 = G_F/\sqrt 2$, and factor
$\eta_1$ takes into account the hard gluon exchanges. Since
\begin{equation}
M_{12}-\frac{i}{2}\Gamma_{12} = <K^0\mid H^{eff}\mid \bar K^0 >
/(2m_K)
\label{74}
\end{equation}
(here $H^{eff} = -{\cal L}_{\Delta s=2}^{eff}) \;\;  \nonumber $
we should calculate the matrix element of
the product of two $V-A$ quark currents between $\bar K^0$ and
$K^0$ states. Using the vacuum insertion we obtain:
\begin{eqnarray}
\langle K^0
\mid\bar s\gamma_\alpha(1+\gamma_5)d \bar s
\gamma_\alpha(1+\gamma_5)d \mid \bar K^0 \rangle = ~~~~~~~~~~~~~~~~~~~~~~~~~ \nonumber \\
= \frac{8}{3}
B_K \langle K^0 \mid\bar s \gamma_\alpha(1+\gamma_s)d \mid 0
\rangle \cdot
\langle 0 \mid \bar s
\gamma_\alpha(1+\gamma_5) d\mid \bar K^0
\rangle = -\frac{8}{3} B_K f_K^2 m_K^2 \;\; ,
\label{75}
\end{eqnarray}

where $B_K =1$ if the vacuum insertion saturates this matrix
element.

From (\ref{62}) we obtain:
\begin{equation}
m_S - m_L -
\frac{i}{2}(\Gamma_S - \Gamma_L) = 2[Re M_{12} -
\frac{i}{2}\Gamma_{12}] \;\; ,
\label{76}
\end{equation}
where $S$ and $L$
are the abbreviations for $K_S$ and $K_L$, short and long-lived
neutral $K$-mesons respectively. For the difference of masses
we get:
\begin{equation}
m_L -m_S \equiv \Delta m_{LS} =
\frac{G_F^2 B_K f_K^2 m_K}{6\pi^2} \eta_1 m_c^2 |V_{cs}^2
V_{cd}^{*^2}| \;\; .
\label{77}
\end{equation}

Constant $f_K$ is known from $K \to l\nu$ decays, $f_K = 160$ MeV.
Gluon dressing of the box diagrams in 4 quark model in the leading
logarithmic (LO) approximation gives
$\eta_1^{LO} = 0.6$. It appears that the subleading logarithms are
numerically very important, $\eta_1^{NLO} = 1.3 \pm
0.2$, the number which we will use in our estimates. We take $B_K
= 0.8\pm 0.1$ assuming that the vacuum insertion is good
numerically, though the smaller values of $B_K$ can be found in
literature as well.

Experimentally the difference of masses is:
\begin{equation}
\Delta m_{LS}^{\rm
exp} = 0.5303(9) \cdot 10^{10} \; {\rm sec}^{-1} \;\; .
\label{78}
\end{equation}

Substituting the numbers  we get:
\begin{equation}
\frac{\Delta
m_{LS}^{\rm theor}}{\Delta m_{LS}^{\rm exp}} = 0.5 \pm 0.2 \;\; ,
\label{79}
\end{equation}
and we almost get an experimental number from the
short-distance contribution
described by the box diagram with
$c$-quarks. As $V_{cs}$ and $V_{cd}$ are already known nothing new
for CKM matrix elements can be extracted from $\Delta m_{LS}$.

However, the very existence of a charm quark and its mass below 2 GeV
were predicted BEFORE 1974 November revolution ($J/\Psi (c\bar c)$ discovery,
$M_{J/\Psi}$ = 3.1 GeV)
from the value of $\Delta m_{LS}$.

Concerning the neutral kaon decays we have:
\begin{equation}
\Gamma_S -\Gamma_L =
2\Gamma_{12} \approx \Gamma_S = 1.1 \cdot 10^{10} \; {\rm
sec}^{-1} \;\; (\Delta m_{LS} \approx \Gamma_S/2) \;\; ,
\label{80}
\end{equation}
since $\Gamma_L \ll \Gamma_S$, $\Gamma_L = 2 \cdot
10^7 \; {\rm sec}^{-1}$.
$K_S$ rapidly decays to two pions which have  CP$=+1$.

$D^0 - \bar D^0$ mixing is established but it is very small:
$\Delta m/\Gamma, \Delta \Gamma/\Gamma \sim10^{-3}$. One of the reasons is the
absence of Cabbibo suppression of $c$-quark decay, while $D^0 - \bar D^0$
transition amplitude is proportional to $\sin^2\theta_c$.

\subsection{CPV in $\boldmath K^0 - \bar K^0: K_L \to 2\pi
\; , \;\; \varepsilon_K$-hyperbola}

CPV in $K^0 - \bar K^0$ mixing is proportional to the deviation of
$\mid q/p\mid$ from one; so let us calculate this ratio
 taking into account that $\Gamma_{12}$ is real, while
$M_{12}$ is mostly real:
\begin{equation}
\frac{q}{p} = 1-\frac{i Im
M_{12}}{M_{12}-\frac{i}{2}\Gamma_{12}} = 1+ \frac{2i Im
M_{12}}{m_L - m_S +\frac{i}{2}\Gamma_S} \;\; .
\label{81}
\end{equation}

In this way for quantity $\tilde\varepsilon$
 we obtain:
\begin{equation}
\tilde\varepsilon = -\frac{i Im
M_{12}}{\Delta m_{LS} + \frac{i}{2}\Gamma_S} \;\; .
\label{82}
\end{equation}

Branching of CP-violating $K_L \to 2\pi$ decay equals:
\begin{eqnarray}
Br (K_L
\to 2\pi^0) +  Br (K_L \to \pi^+ \pi^-) = \frac{\Gamma(K_L \to
2\pi)}{\Gamma_{K_L}} = \frac{\Gamma_{K_L \to 2\pi}}{\Gamma_{K_S
\to 2\pi}} \frac{\Gamma(K_S)}{\Gamma(K_L)} = \nonumber \\
 =
\frac{\mid\eta_{00}\mid^2 \Gamma(K_S \to 2\pi^0) +
\mid\eta_{+-}\mid^2 \Gamma(K_S \to \pi^+ \pi^-)}{\Gamma(K_S \to
2\pi^0) +\Gamma(K_S\to \pi^+ \pi^-)}
\frac{\Gamma(K_S)}{\Gamma(K_L)} \approx  \nonumber \\
 \approx
\mid\eta_{00}\mid^2 \frac{\Gamma(K_S)}{\Gamma(K_L)} \approx
\mid\tilde\varepsilon\mid^2 \frac{\Gamma(K_S)}{\Gamma(K_L)}
\approx \mid\tilde\varepsilon\mid^2 \frac{5.12(2) \cdot
10^{-8}\; {\rm sec}}{0.895(0.3) \cdot 10^{-10}\; {\rm sec}} \approx \nonumber \\
 \approx 572 \mid\tilde\varepsilon \mid^2 = 2.83(1)
\cdot 10^{-3} \;\; ,
\label{83}
\end{eqnarray}

where the last number is the sum of $K_L
\to \pi^+ \pi^-$ and $K_L \to \pi^0 \pi^0$ branching ratios.
In this way the experimental value of $\mid\tilde\varepsilon\mid$
is determined, and for a theoretical result
 we should have:
\begin{equation}
\mid\tilde\varepsilon\mid = \frac{\mid Im
M_{12}\mid}{\sqrt 2 \Delta m_{LS}} = 2.22 \cdot 10^{-3}.
\label{84}
\end{equation}

As we have already demonstrated, ($tt$) box gives the main contribution to $Im M_{12}$.
It was calculated for the first time explicitly not supposing that $m_t\ll m_W$ in 1980 \cite{17}:
\begin{eqnarray}
Im M_{12} = -\frac{G_F^2
B_K f_K^2 m_K}{12\pi^2} m_t^2 \eta_2 Im(V_{ts}^2 V_{td}^{*^2})
\times I(\xi) \;\; , \nonumber \\
I(\xi) = \left\{\frac{\xi^2 -11\xi
+4}{4(\xi -1)^2} -\frac{3\xi^2 \ln\xi}{2(1-\xi)^3} \right\} \; ,
\;\; \xi =\left(\frac{m_t}{m_W}\right)^2 \;\; ,
\label{85}
\end{eqnarray}
where factor $\eta_2$ takes into account the gluon exchanges
in the box diagram with ($tt$) quarks and in the leading logarithmic
approximation it equals
$\eta_2^{LO}=0.6$. This factor is not changed substantially by
subleading logs: $\eta_2^{NLO} = 0.57(1)$.

Let us present the numerical values for the expression in figure
brackets  for several values of the top quark mass:
\begin{equation}
\left\{ \;\;
\right\}
=
\begin{array}{cl} 1 \; , & m_t =0 \; , \;\; \xi =0
\\ 0.55 \; , & \xi =4.7 \; , \;\; {\rm which ~~ corresponds ~~ to}
\;\; m_t = 175 \; {\rm GeV}
\\ 0.25 \; , & m_t = \xi = \infty \end{array}
\label{86}
\end{equation}

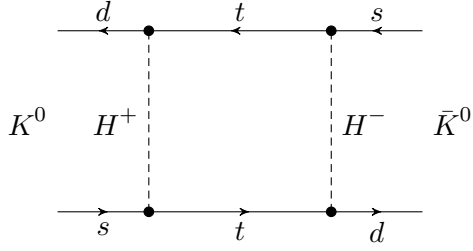
\begin{figure}[t]
  \centering
  \begin{tikzpicture}[scale=1.2]
    \tikzset{every node/.append style={font=\normalsize}}
      \coordinate (A) at (0,0);
      \coordinate (B) at (1,0);
      \coordinate (C) at (3,0);
      \coordinate (D) at (4,0);
      \coordinate (E) at (0,2);
      \coordinate (F) at (1,2);
      \coordinate (G) at (3,2);
      \coordinate (H) at (4,2);
      \coordinate (S1) at (0,1);
      \coordinate (S2) at (4,1);
      \draw[electron] (A) to node[below]{$s$} (B);
      \draw[electron] (B) to node[below]{$t$} (C);
      \draw[electron] (C) to node[below]{$d$} (D);
      \draw[electron] (F) to node[above]{$ d$} (E);
      \draw[electron] (G) to node[above]{$ t$} (F);
      \draw[electron] (H) to node[above]{$ s$} (G);
       \draw[scalar] (B) to node[left]{$H^{+}$} (F);
       \draw[scalar] (C) to node[right]{$H^{-}$} (G);
      \draw (S1) node[left]{$K^{0}$};
      \draw (S2) node[right]{$\bar K^{0}$};
      \filldraw (B) circle (1.5pt);
      \filldraw (C) circle (1.5pt);
      \filldraw (F) circle (1.5pt);
      \filldraw (G) circle (1.5pt);
     \end{tikzpicture}
  \caption{The diagram which dominates in the limit $m_t \gg m_W$.}
  \label{fig:hh}
\end{figure}

It is clearly seen that the top contribution to the box diagram is not decoupled:
it does not vanish in the limit $m_t \to \infty$.
One can easily get where this enhanced at $m_t \to\infty$
behaviour originates by estimating the box diagram in
't~Hooft-Feynman gauge. In the limit $m_t \gg m_W$ the diagram
with two charged higgs exchanges dominates (see \Figure\ref{fig:hh}), since
each vertex of higgs boson emission is proportional to $m_t$.

For the factor which multiplies the four-quark operator from this diagram
we get:
\begin{equation}
\sim (\frac{m_t}{v})^4 \int\frac{d^4
p}{(p^2 -M_W^2)^2} \left[\frac{\hat p}{p^2 -m_t^2}\right]^2 \sim
(\frac{ m_t}{v})^4 \frac{1}{ m_t^2} = G_F^2  m_t^2 \;\; ,
\label{87}
\end{equation}
where $v$ is the higgs boson expectation value.  No decoupling!

Substituting the numbers we obtain:
\begin{equation}
\eta(1-\rho) = 0.47(5) \;\; ,
\label{88}
\end{equation}
where 10\% uncertainty in
the value of $B_K = 0.8 \pm 0.1$ dominates in the error. Taking into
account ($ct$) and ($cc$) boxes we get the following equation:
\begin{equation}
\eta(1.4 -\rho) = 0.47(5) \;\; -
\label{89}
\end{equation}
 hyperbola on ($\rho, \eta$) plane.

\bigskip

Why is $\varepsilon_K$ so small?  We have the following estimate for
$\varepsilon_K$:
\begin{equation}
\varepsilon_K \sim \frac{m_t^2 \lambda^{10} \eta(1-\rho)}{m_c^2
\lambda^2} \;\; .
\label{90}
\end{equation}

It means that $\varepsilon_K$ is small not because CKM phase is
small, but because $2\times 2$ part of CKM matrix which describes the
mixing of the first two generations is almost unitary and the
third generation almost decouples. We are lucky that the top quark
is so heavy; for $m_t \sim 10$ GeV CPV  would not have been discovered in
1964.

\bigskip

\section{Direct CPV}
\subsection{Direct CPV in $\boldmath K$ decays,
$\boldmath\varepsilon^\prime \neq 0 \; (\mid\frac{\bar
A}{A}\mid \neq 1)$}

Let us consider the neutral kaon decays into two pions. It is
convenient to deal with the amplitudes of the decays into the
states with a definite isospin:
\begin{equation}
A(K^0 \to \pi^+ \pi^-) =
\frac{a_2}{\sqrt 3} e^{i\xi_2}e^{i\delta_2} + \frac{a_0}{\sqrt 3}
\sqrt{2} e^{i\xi_0} e^{i\delta_0} \;\; ,
\label{91}
\end{equation}
\begin{equation}
A(\bar
K^0 \to \pi^+ \pi^-) = \frac{a_2}{\sqrt 3}
e^{-i\xi_2}e^{i\delta_2} + \frac{a_0}{\sqrt 3} \sqrt{2}
e^{-i\xi_0} e^{i\delta_0} \;\; ,
\label{92}
\end{equation}
\begin{equation}
A(K^0 \to \pi^0
\pi^0) = \sqrt{\frac{2}{3}} a_2 e^{i\xi_2}e^{i\delta_2} -
\frac{a_0}{\sqrt 3} e^{i\xi_0} e^{i\delta_0} \;\; ,
\label{93}
\end{equation}
\begin{equation}
A(\bar K^0 \to \pi^0 \pi^0) = \sqrt{\frac{2}{3}} a_2
e^{-i\xi_2}e^{i\delta_2} - \frac{a_0}{\sqrt 3} e^{-i\xi_0}
e^{i\delta_0} \;\; ,
\label{94}
\end{equation}
where ``2'' and ``0'' are the
values of ($\pi\pi$) isospin, $\xi_{2,0}$ are the weak phases
which originate from CKM matrix and $\delta_{2,0}$ are the strong
phases of $\pi\pi$-rescattering. If the only quark diagram
responsible for $K\to 2\pi$ decays were the charged current tree
diagram which describes $s\to u\bar u d$ transition through
$W$-boson exchange, then the weak phases would be zero and it would be
no CPV in the decay amplitudes (the so-called direct CPV). All CPV
would originate from $K^0 - \bar K^0$ mixing. Such indirect CPV
was called superweak (L.Wolfenstein, 1964).

\begin{center}
\begin{figure}[h]
\hspace{45mm}
\includegraphics[width=.4\textwidth]{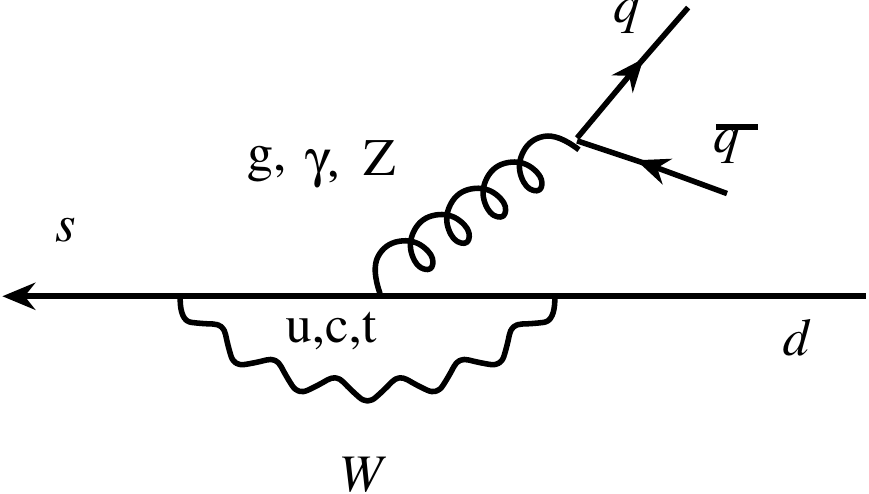}
\caption{The penguin diagrams contributing to kaon decays.}
\label{fig:strpen}
\end{figure}

\end{center}

However, in Standard Model the CKM phase
penetrates into the amplitudes of $K\to 2\pi$ decays through the
so-called  ``penguin'' diagrams shown in \Figure\ref{fig:strpen} and $\xi_{0}$
and $\xi_{2}$ are
nonzero leading to direct CPV as well.

From (\ref{91}) and (\ref{92}) we get:
\begin{equation}
\Gamma(K^0 \to \pi^+ \pi^-)
-\Gamma(\bar K^0 \to \pi^+ \pi^-) = -4\frac{\sqrt 2}{3} a_0 a_2
\sin(\xi_2 -\xi_0)\sin(\delta_2 -\delta_0) \;\; ,
\label{95}
\end{equation}
so for direct CPV to occur through the difference of $K^0$ and $\bar K^0$
widths at least two decay amplitudes with different CKM and strong
phases should exist.

In the decays of $K_L$ and $K_S$ mesons the violation of CP occurs due
to that in mixing (indirect CPV) and in decay amplitudes of $K^0$
and $\bar K^0$ (direct CPV). The first effect is taken into
account in the expression for $K_L$ and $K_S$ eigenvectors through
$K^0$ and $\bar K^0$:
\begin{equation}
K_S = \frac{K^0 +\bar K^0}{\sqrt 2} + \tilde\varepsilon
\frac{K^0 - \bar K^0}{\sqrt 2} \;\; ,
\label{96}
\end{equation}
\begin{equation}
K_L =
\frac{K^0 -\bar K^0}{\sqrt 2} + \tilde\varepsilon \frac{K^0 + \bar
K^0}{\sqrt 2} \;\; ,
\label{97}
\end{equation}
where we neglect $\sim
\tilde\varepsilon^2$ terms.
For the amplitudes of $K_L$ and $K_S$ decays
into $\pi^+ \pi^-$  we obtain:
\begin{eqnarray}
A(K_L \to\pi^+ \pi^-) = \frac{1}{\sqrt 2} \left[
\frac{a_2}{\sqrt 3} e^{i\delta_2} 2i\sin\xi_2 + \frac{a_0}{\sqrt
3} \sqrt 2 e^{i\delta_0} 2i\sin\xi_0 \right] + \nonumber \\
+\frac{\tilde\varepsilon}{\sqrt 2} \left[ \frac{a_2}{\sqrt 3}
e^{i\delta_2} 2\cos\xi_2 + \frac{a_0}{\sqrt 3} \sqrt 2
e^{i\delta_0} 2\cos\xi_0 \right] \;\; ,
\label{98}
\end{eqnarray}
\begin{equation}
A(K_S
\to\pi^+ \pi^-) = \frac{1}{\sqrt 2} \left[ \frac{a_2}{\sqrt 3}
e^{i\delta_2} 2\cos\xi_2 + \frac{a_0}{\sqrt 3} \sqrt 2
e^{i\delta_0} 2\cos\xi_0 \right] \;\; ,
\label{99}
\end{equation}
where in
the last equation we omit the terms which are proportional to the
product of two small factors, $\tilde\varepsilon$ and $\sin\xi_{0,2}$.
For the ratio of these amplitudes we get:
\begin{equation}
\eta_{+-} \equiv
\frac{A(K_L \to\pi^+\pi^-)}{A(K_S \to\pi^+ \pi^-)} =
\tilde\varepsilon + i\frac{\sin\xi_0}{\cos\xi_0} +\frac{i
e^{i(\delta_2 - \delta_0)}}{\sqrt 2} \frac{a_2 \cos\xi_2}{a_0
\cos\xi_0} \left[ \frac{\sin\xi_2}{\cos\xi_2} -
\frac{\sin\xi_0}{\cos\xi_0}\right] \;\; ,
\label{100}
\end{equation}
where we
neglect the terms of the order of $(a_2/a_0)^2 \sin\xi_{0,2}$
because from the
$\Delta I = 1/2$ rule in $K$-meson decays it is
known that $a_2/a_0 \approx 1/22$.

The analogous treatment of $K_{L,S} \to \pi^0 \pi^0$ decay
amplitudes leads to:
\begin{equation}
\eta_{00} \equiv \frac{A(K_L
\to\pi^0\pi^0)}{A(K_S \to \pi^0 \pi^0)} = \tilde\varepsilon +
i\frac{\sin\xi_0}{\cos\xi_0} - ie^{i(\delta_2 - \delta_0)} \sqrt 2
\frac{a_2 \cos\xi_2}{a_0 \cos\xi_0}
\left[\frac{\sin\xi_2}{\cos\xi_2} - \frac{\sin\xi_0}{\cos\xi_0}
\right] \;\; .
\label{101}
\end{equation}

The difference of $\eta_{\pm}$ and $\eta_{00}$ is proportional to
$\varepsilon^\prime$:
\begin{eqnarray}
\varepsilon^\prime \equiv \frac{i}{\sqrt
2} e^{i(\delta_2 - \delta_0)} \frac{a_2 \cos\xi_2}{a_0 \cos\xi_0}
\left[\frac{\sin\xi_2}{\cos\xi_2} - \frac{\sin\xi_0}{\cos\xi_0}
\right] = \\
= \frac{i}{\sqrt 2} e^{i(\delta_2 -
\delta_0)} \frac{Re A_2}{Re A_0} \left[ \frac{Im A_2}{Re A_2} -
\frac{Im A_0}{Re A_0} \right] = \frac{i}{\sqrt 2} e^{i(\delta_2
-\delta_0)} \frac{1}{Re A_0} \left[ Im A_2 - \frac{1}{22} Im A_0
\right] \; , \nonumber
\label{102}
\end{eqnarray}
where $A_{2,0} \equiv e^{i\xi_{2,0}} a_{2,0}$.

Introducing quantity $\varepsilon$ according to the standard
definition
\begin{equation}
\varepsilon = \tilde\varepsilon + i \frac{Im A_0}{Re
A_0} \;\; ,
\label{103}
\end{equation}
we obtain:
\begin{equation}
\eta_{+-} = \varepsilon +
\varepsilon^\prime \; , \;\; \eta_{00} = \varepsilon -
2\varepsilon^\prime \;\; .
\label{104}
\end{equation}

The double ratio $\eta_{+-}/\eta_{00}$ was measured in the
experiment and its difference from 1 demonstrates direct CPV in
kaon decays:
\begin{equation}
\left(\frac{\varepsilon^\prime}{\varepsilon}\right)^{\rm exp} =
(1.67 \pm 0.23) \cdot 10^{-3} \;\; .
\label{105}
\end{equation}

The smallness of this ratio is due to (1) the smallness of the
phases produced by the penguin diagrams  and (2)
smallness of the ratio  $a_2/a_0 \approx Re A_2/Re A_0$.

Let us estimate the numerical value of $\varepsilon^\prime$. The penguin
diagram with the gluon exchange generates $K\to 2\pi$ transition
with $\Delta I = 1/2$; those with $\gamma$- and $Z$-exchanges
contribute to $\Delta I = 3/2$ transitions as well. The
contribution of electroweak penguins being smaller by the ratio of
squares of coupling constants is enhanced by the factor $Re A_0/Re
A_2 = 22$, see the last part in equation for $\varepsilon^\prime $.
As a result the partial
compensation of QCD and electroweak penguins occurs. In order to
obtain an order of magnitude estimate let us take into account
only QCD penguins. We obtain the following
estimate for the sum of the loops with $t$- and $c$-quarks:
\begin{equation}
\mid\varepsilon^\prime\mid \approx \frac{1}{22\sqrt{2}}
\frac{\sin\xi_0}{\cos\xi_0} =
\frac{1}{22\sqrt{2}}\frac{\alpha_s(m_c)}{12\pi}
\ln(\frac{m_t}{m_c})^2 A^2 \lambda^4 \eta \approx
2*10^{-5}
\frac{\alpha_s(m_c)}{12\pi} \ln (\frac{m_t}{m_c})^2 \;\; .
\label{106}
\end{equation}

Taking into account that $\mid\varepsilon\mid \approx 2.4 \cdot
10^{-3}$ we see that the smallness of the ratio of
$\varepsilon^\prime/\varepsilon$ can be readily understood.

In order to make an accurate calculation of
$\varepsilon^\prime/\varepsilon$ one should know the matrix
elements of the quark operators between $K$-meson and two
$\pi$-mesons. Unfortunately at low energies our knowledge of
QCD is not enough for such a calculation.
That is why a horizontal strip  to which
an apex of the unitarity triangle should belong according to equation for $\varepsilon^\prime/\varepsilon$
has too large width and usually is not
shown. Nevertheless we have discussed direct CPV since
it will be important for $B$ and $D$-mesons.

\subsection{Direct CP asymmetries in $D^0 (\bar D^0)\rightarrow \pi^+\pi^-, \;\; K^+K^-$}

The following result was reported by LHCb collaboration in 2019 \cite{15}:
\begin{equation}
\Delta A_{CP} = A_{CP}( K^+K^-) - A_{CP}( \pi^+\pi^-) = (-15.4\pm2.9)\times10^{-4},
\label{107}
\end{equation}
where CP asymmetry is defined as
\begin{equation}
A_{CP}(f) = \frac{\Gamma(D^0\rightarrow f) - \Gamma(\bar D^0\rightarrow f)}
{\Gamma(D^0\rightarrow f) + \Gamma(\bar D^0\rightarrow f)}.
\label{108}
\end{equation}

To distinguish $D_0$ from $\bar D_0$ the tagging by the charge of pions in
$D^{*+}\rightarrow D^0\pi^+,
D^{*-}\rightarrow \bar D^0\pi^-$ decays and by the charge of muon in semileptonic
$\bar B \to D^0 \mu^- \bar \nu_{\mu} X$ decays has been performed.

\begin{figure}[h]
  \centering
  \begin{tikzpicture}[scale=0.8 ]
    \tikzset{every node/.append style={font=\normalsize}}
    %
    %
    \coordinate (A) at (0,0);
    \coordinate (D) at (4,0);
    \coordinate (E) at (0,2);
    \coordinate (H) at (4,2);
    \coordinate (S1) at (0,1);
    \coordinate (S2) at (4,1);
    \coordinate (S3) at (4,3.5);
    \coordinate (C2) at (2,2);
    \coordinate (X) at (2.5,3.5);
    \coordinate (Y) at (4,4.2);
    \coordinate (Z) at (4,2.8);
    \draw[electron] (A) node[above]{$u$} to (D) node[above]{$u$};
    \draw[electron] (C2) to (E) node[below]{$c$};
    \draw[electron] (H) node[below]{$d$} to (C2);
    \draw[electron] (X) to (Y) node[right]{$d$};
    \draw[electron] (Z) node[right]{$u$} to (X);
    \draw[photon] (C2) to node[left]{$W$} (X);
    \draw (S1) node[left]{$\bar D^0$};
    \draw (S2) node[right]{$ $};
    \filldraw (X) circle (1.5pt);
    \filldraw (C2) circle (1.5pt);
  \end{tikzpicture}
  \hspace{15mm}
  \begin{tikzpicture}[scale=0.8]
    \tikzset{every node/.append style={font=\normalsize}}
    %
    %
    \coordinate (A) at (0,0);
    \coordinate (D) at (4,0);
    \coordinate (E) at (0,2);
    \coordinate (F) at (1,2);
    \coordinate (G) at (3,2);
    \coordinate (H) at (4,2);
    \coordinate (S1) at (0,1);
    \coordinate (S2) at (4,1);
    \coordinate (S3) at (4,3.5);
    \coordinate (C2) at (2,2);
    \coordinate (X) at (2.5,3.5);
    \coordinate (Y) at (4,4.2);
    \coordinate (Z) at (4,2.8);
    \draw[electron] (A) node[above]{$u$} to (D) node[above]{$u$};
    \draw[electron] (F) to (E) node[below]{$c$};
    \draw[electron] (C2) to node[above]{$b$} (F);
    \draw[electron] (G) to node[above]{$b$} (C2);
    \draw[electron] (H) node[below]{$u$} to (G);
    \draw[electron] (X) to (Y) node[right]{$d$};
    \draw[electron] (Z) node[right]{$d$} to (X);
    \draw[gluon] (C2) to node[left]{$g$} (X);
    \draw[photon] (F) ++ (0,-1.6pt) arc (188:360:1 and 0.76) ;
    \draw (S1) node[left]{$\bar D^0$};
    \draw (S2) node[right]{$ $};
    \filldraw (X) circle (1.5pt);
    \filldraw (C2) circle (1.5pt);
    \filldraw (F) circle (1.5pt);
    \filldraw (G) circle (1.5pt);
  \end{tikzpicture}
  \hfill
 \caption{The diagrams responsible for $\bar D^0 \to \pi^+\pi^-$ decay.}
  \label{fig:dd}
\end{figure}
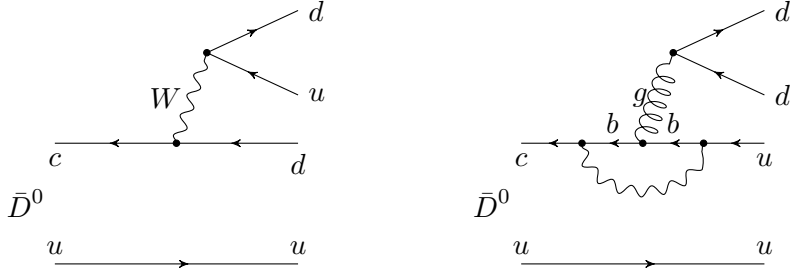

The interference of tree and penguin amplitudes shown in \Figure\ref{fig:dd} leads to
CP asymmetry:
\begin{equation}
 A(\bar D)=e^{i\delta}TV_{cd}V^*_{ud} - P V_{cb}|V_{ub}|e^{i\gamma},
 \label{109}
 \end{equation}
\begin{equation}
A(D)=e^{i\delta}TV^*_{cd}V_{ud} - P V^*_{cb}|V_{ub}|e^{-i\gamma},
\label{110}
\end{equation}
\begin{equation}
A_{CP}( \pi^+\pi^-) = \frac{4TPV_{cd}V^*_{ud} |V_{ub}|V^*_{cb}\sin(\delta)\sin(\gamma)}
{2T^2|V_{cd}V_{ud}|^2}.
\label{111}
\end{equation}

In the limit of $U$-spin $(d\leftrightarrow s)$ symmetry $ A_{CP}( K^+K^-)=
- A_{CP}( \pi^+\pi^-)$, and sign ``-" comes from $V_{cd} = - V_{us}$.
Thus we get:
\begin{equation}
|\Delta A_{CP}| = 4|P/TA^2\lambda^4\sqrt{\rho^2+\eta^2}\sin(\delta)\sin(\gamma)|
\approx |25 \sin(\delta) P/T|\times 10^{-4},
\label{112}
\end{equation}
and to reproduce an experimental result strong interactions phase $\delta$
should be big and penguin
amplitude should be of the order of the tree one.

The reason for the small value of CPV asymmetry in charm is the same as in $K$-
mesons: $2\times 2$ part of CKM matrix which describes mixing of the first
and second generations is almost unitary.
The absence of $\Delta I=1/2$ amplitude
enhancement  makes direct CPV asymmetry in case of $D$ decays larger than in kaon decays.

When the third generation is involved CPV can be big.

\begin{center}
\begin{figure}[h]

\includegraphics[width=.8\textwidth]{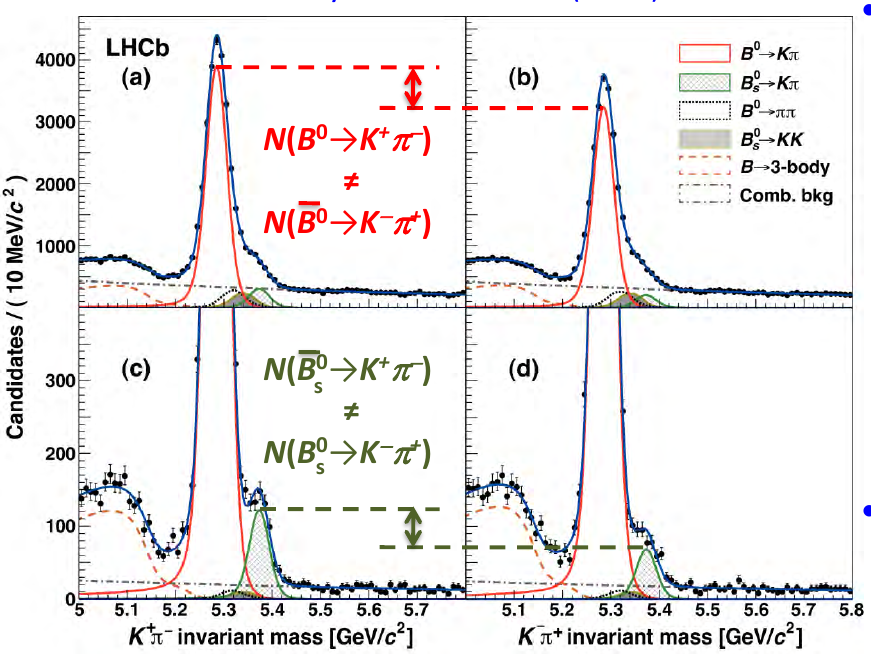}

\caption{Direct CPV in $B^0(B_s^0)\to K\pi$ decays.}
\label{fig:kpi}
\end{figure}
\end{center}

\subsection{25 \% direct CP asymmetry in $B_s$ decay}
 While direct CPV in kaons and $D$-mesons is very small it is sometimes huge in B-mesons, see \Figure\ref{fig:kpi} \cite{18}.

The diagrams shown in \Figure\ref{fig:bskpi} describe $B_s \to K^-\pi^+$ decay.

\begin{figure}[h]
\centering
  \begin{tikzpicture}[scale=0.7]
    \tikzset{every node/.append style={font=\normalsize}}
    %
    %
    \coordinate (A) at (0,0);
    \coordinate (D) at (4,0);
    \coordinate (E) at (0,2);
    \coordinate (H) at (4,2);
    \coordinate (S1) at (0,1);
    \coordinate (S2) at (4,1);
    \coordinate (S3) at (4,3.5);
    \coordinate (C2) at (2,2);
    \coordinate (X) at (2.5,3.5);
    \coordinate (Y) at (4,4.2);
    \coordinate (Z) at (4,2.8);
    \draw[electron] (A) node[above]{$s$} to (D) node[above]{$s$};
    \draw[electron] (C2) to (E) node[below]{$b$};
    \draw[electron] (H) node[below]{$u$} to (C2);
    \draw[electron] (X) to (Y) node[right]{$u$};
    \draw[electron] (Z) node[right]{$d$} to (X);
    \draw[photon] (C2) to node[left]{$W$} (X);
    \draw (S1) node[left]{$B_s$};
    \draw (S2) node[right]{$ $};
    \filldraw (X) circle (1.5pt);
    \filldraw (C2) circle (1.5pt);
  \end{tikzpicture}
  \hspace{15mm}
  \begin{tikzpicture}[scale=0.7]
    \tikzset{every node/.append style={font=\normalsize}}
    %
    %
    \coordinate (A) at (0,0);
    \coordinate (D) at (4,0);
    \coordinate (E) at (0,2);
    \coordinate (F) at (1,2);
    \coordinate (G) at (3,2);
    \coordinate (H) at (4,2);
    \coordinate (S1) at (0,1);
    \coordinate (S2) at (4,1);
    \coordinate (S3) at (4,3.5);
    \coordinate (C2) at (2,2);
    \coordinate (X) at (2.5,3.5);
    \coordinate (Y) at (4,4.2);
    \coordinate (Z) at (4,2.8);
    \draw[electron] (A) node[above]{$s$} to (D) node[above]{$s$};
    \draw[electron] (F) to (E) node[below]{$b$};
    \draw[electron] (C2) to node[above]{$c$} (F);
    \draw[electron] (G) to node[above]{$c$} (C2);
    \draw[electron] (H) node[below]{$d$} to (G);
    \draw[electron] (X) to (Y) node[right]{$u$};
    \draw[electron] (Z) node[right]{$u$} to (X);
    \draw[gluon] (C2) to node[left]{$g$} (X);
    \draw[photon] (F) ++ (0,-1.6pt) arc (188:360:1 and 0.76) ;
    \draw (S1) node[left]{$B_s$};
    \draw (S2) node[right]{$ $};
    \filldraw (X) circle (1.5pt);
    \filldraw (C2) circle (1.5pt);
    \filldraw (F) circle (1.5pt);
    \filldraw (G) circle (1.5pt);
  \end{tikzpicture}
  \hfill
 \caption{$B_s \to K^-\pi^+$ decay.}
  \label{fig:bskpi}
\end{figure}

\begin{equation}
A( B_s \longrightarrow K^-\pi^+)= T_s V^*_{ub}V_{ud} + P_s e^{i\delta}
 V^*_{cb}V_{cd},
 \label{113}
 \end{equation}
\begin{equation}
 A(\bar B_s \longrightarrow K^+\pi^-)= T_s V_{ub}V^*_{ud} + P_s e^{i\delta}
 V_{cb}V^*_{cd},
 \label{114}
 \end{equation}
where $\delta$ is strong phase; CKM phase is contained in
  $V_{ub}=-e^{-i\gamma}|V_{ub}|$.
\begin{eqnarray}
A_{CP}( B_s \longrightarrow K^-\pi^+)= \frac{|A(\bar B_s)|^2-
 |A(B_s)|^2 }{|A(\bar B_s)|^2+
 |A(B_s)|^2 }= \\
=\frac{4T_s P_s V^*_{ud}V_{cb}V^*_{cd}|V_{ub}|\sin(\delta)\sin
 (\gamma)}{2T_s^2|V_{ub}V_{ud}|^2+2P_s^2|V_{cb}V_{cd}|^2-4P_s T_s V^*_{ud}
 V_{cb}V^*_{cd}|V_{ub}|\cos(\delta)\cos(\gamma)}.\nonumber
 \label{115}
 \end{eqnarray}

 CKM factors in the nominator and denominator
 are of the order of $\lambda^6$ and there is no CKM suppression
 of $A_{CP}(B_s)$. Since asymmetry is big, $P_s/T_s$ is not that small.

 \begin{figure}[h]
  \centering
  \begin{tikzpicture}[scale=0.7]
    \tikzset{every node/.append style={font=\normalsize}}
    %
    %
    \coordinate (A) at (0,0);
    \coordinate (D) at (4,0);
    \coordinate (E) at (0,2);
    \coordinate (H) at (4,2);
    \coordinate (S1) at (0,1);
    \coordinate (S2) at (4,1);
    \coordinate (S3) at (4,3.5);
    \coordinate (C2) at (2,2);
    \coordinate (X) at (2.5,3.5);
    \coordinate (Y) at (4,4.2);
    \coordinate (Z) at (4,2.8);
    \draw[electron] (A) node[above]{$d$} to (D) node[above]{$d$};
    \draw[electron] (C2) to (E) node[below]{$b$};
    \draw[electron] (H) node[below]{$u$} to (C2);
    \draw[electron] (X) to (Y) node[right]{$u$};
    \draw[electron] (Z) node[right]{$s$} to (X);
    \draw[photon] (C2) to node[left]{$W$} (X);
    \draw (S1) node[left]{$B^{0}$};
    \draw (S2) node[right]{$ $};
    \filldraw (X) circle (1.5pt);
    \filldraw (C2) circle (1.5pt);
  \end{tikzpicture}
  \hspace{45mm}
  \begin{tikzpicture}[scale=0.7]
    \tikzset{every node/.append style={font=\normalsize}}
    %
    %
    \coordinate (A) at (0,0);
    \coordinate (D) at (4,0);
    \coordinate (E) at (0,2);
    \coordinate (F) at (1,2);
    \coordinate (G) at (3,2);
    \coordinate (H) at (4,2);
    \coordinate (S1) at (0,1);
    \coordinate (S2) at (4,1);
    \coordinate (S3) at (4,3.5);
    \coordinate (C2) at (2,2);
    \coordinate (X) at (2.5,3.5);
    \coordinate (Y) at (4,4.2);
    \coordinate (Z) at (4,2.8);
    \draw[electron] (A) to (D) node[above]{$d$};
    \draw[electron] (F) to (E) node[below]{$b$};
    \draw[electron] (C2) to node[above]{$c$} (F);
    \draw[electron] (G) to node[above]{$c$} (C2);
    \draw[electron] (H) node[below]{$s$} to (G);
    \draw[electron] (X) to (Y) node[right]{$u$};
    \draw[electron] (Z) node[right]{$u$} to (X);
    \draw[gluon] (C2) to node[left]{$g$} (X);
    \draw[photon] (F) ++ (0,-1.6pt) arc (188:360:1 and 0.76);
    \draw (S1) node[left]{$B^{0}$};
    \draw (S2) node[right]{$ $};
    \filldraw (X) circle (1.5pt);
    \filldraw (C2) circle (1.5pt);
    \filldraw (F) circle (1.5pt);
    \filldraw (G) circle (1.5pt);
  \end{tikzpicture}
  \hfill
  \caption{$B^0 \to K^+\pi^-$ decay.}
  \label{fig:bkpi}
\end{figure}

Though we cannot compute diagrams in \Figure\ref{fig:bskpi} and \Figure\ref{fig:bkpi}, we can relate them in the
$U$ spin invariance approximation.

\bigskip

Problem 5

  Derive an expression for
  $A_{CP}( B^0 \longrightarrow K^+\pi^-)$ and get the following
  equality:
  \begin{equation}
 A_{CP}( B^0 ) \cdot \Gamma_{B^0\rightarrow K\pi} = -A_{CP}( B_s )
  \cdot \Gamma_{B_s\rightarrow K\pi} \;\; .
  \label{115a}
  \end{equation}

Substituting experimentally measured numbers from RPP (PDG) \cite{19}
  for asymmetries
    $A_{CP}( B^0 )=-0.082(6), \;\;\; A_{CP}( B_s )=0.26(4)$ and
  branching ratios Br($B^0\rightarrow K\pi)=20\cdot10^{-6}$,
  Br($B_s\rightarrow K\pi)=5.7\cdot10^{-6}$ check this equality.

 Smallness of branching ratio of any exclusive decay is the main problem
 in studying CPV in $B$-mesons.

 \subsection{CPV in neutrino oscillations}
  \justifying
  \vspace{-0.5em}

In order to have CPV we need not only CP violating phase but
CP conserving phase as well ($i\Gamma_{12}$ in case of mixing, $\delta_2-\delta_0$ in case of direct CPV in kaon decays).

\bigskip

Problem 6

In case of leptons the flavor mixing is described by the
{\color{purple}P}MNS matrix:
\begin{equation}
\left( \begin{array}{c} \nu_e \\ \nu_\mu \\ \nu_\tau \end{array} \right)  =
\left(\begin{array}{lll} V_{e1} & V_{e2} & V_{e3} \\ V_{\mu1} &
    V_{\mu2} & V_{\mu3} \\ V_{\tau1} & V_{\tau2} & V_{\tau3}
      \end{array} \right)
    \left( \begin{array}{c} \nu_1 \\ \nu_2 \\ \nu_3 \end{array} \right) \;\;.
    \label{116}
    \end{equation}

 CPV means in particular that the probability of $\nu_\mu\longrightarrow \nu_e$
 oscillation $P_{e\mu}$ does not coincide with the probability of
 $\bar\nu_\mu\longrightarrow \bar\nu_e$ oscillation
 $P_{\bar e\bar\mu}$.

 Check that
\begin{equation}
 P_{e\mu} - P_{\bar e\bar\mu} = 4 Im(V_{\mu1}^* V_{e1} V_{\mu2} V_{e2}^*)
 * [\sin(\frac{\Delta m^2_{12}}{2E}x ) + \sin(\frac{\Delta
   m^2_{31}}{2E}x ) + \sin(\frac{\Delta m^2_{23}}{2E}x )].
   \label{117}
   \end{equation}

Just like in kaons CPV is proportional to Jarlskog invariant.

When two neutrinos have equal masses there is no CPV.

 Where is the CP conserving phase in the case of CPV in neutrino oscillations?

 By the way, the driving force for Bruno Pontecorvo to consider neutrino
 oscillations was the observed oscillations of neutral kaons \cite{20}.

\subsection {CPV - absolute notion of a particle }
\begin{equation}
\delta_L = \frac{\Gamma(K_L\rightarrow\pi^-e^+\nu) -
\Gamma(K_L\rightarrow\pi^+e^-\bar\nu)}{ \Gamma(K_L\rightarrow\pi^-e^+\nu) +
\Gamma(K_L\rightarrow\pi^+e^-\bar\nu)}= 2Re \tilde\varepsilon\approx
3.3*10^{-3}.
\label{118}
\end{equation}

\bigskip
 Pions of low energies mostly produce $K^0$ on the Earth, while $\bar K^0$
 on the ``antiEarth'' ($\pi N \to K^0(\Lambda, \Sigma)$; $\pi \bar N \to \bar K^0(\bar\Lambda,
 \bar\Sigma)$). However, in both cases $K_L$ decay (a little bit)
 more often into positrons than into electrons.

 ``The atoms on the Earth contain antipositrons (electrons) - and
 what about your planet?''

In this way the measurements of the probabilities of semileptonic $K_L$
decays allow to decide if the other planet is made from antimatter.

 \bigskip

 Problem 7

   Violation of leptonic (muon and electron) numbers due to neutrino
   mixing. Estimate the branching ratio of the $\mu\longrightarrow e \gamma$
   decay, which occurs in the Standard Model due to the analog of the
   penguin diagram from \Figure\ref{fig:strpen} without splitting of the photon.

\bigskip

\section{Constraints on the unitarity triangle}
\subsection{Parameters of CKM matrix}
 \begin{center}
\begin{figure}[h]
\hspace{20mm}
\includegraphics[width=.7\textwidth]{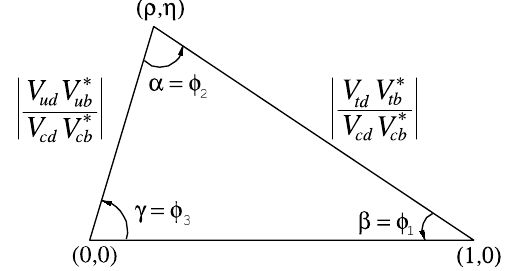}

\end{figure}

  \end{center}
\begin{center}
\begin{figure}[h]
\hspace{25mm}
    \includegraphics[width=.6\textwidth]{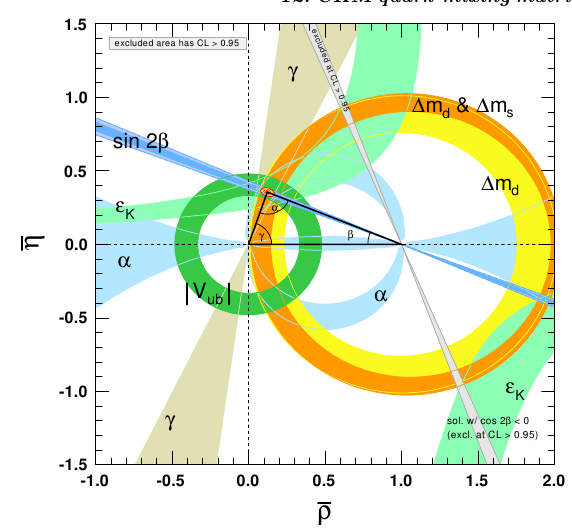}
\caption{Constraints on the apex of the unitarity triangle.}
 \label{fig:utc}
\end{figure}

\end{center}

Four quantities are needed to specify CKM matrix: $s_{12}, s_{13},
s_{23}$ and $\delta$, or $\lambda, A, \rho, \eta$.
The areas shaded in \Figure\ref{fig:utc} \cite{31} show the domains of $\bar \rho$ and $\bar \eta$
allowed at 95\% C.L. by different measurements ($\bar \rho \equiv \rho (1-\lambda^2/2), \;\;
\bar \eta \equiv \eta (1-\lambda^2/2)$).

\subsection{$V_{cd}, V_{cb}, V_{ub}$}

\bigskip
The precise value of $V_{us}$ follows from the
extrapolation of the formfactor of $K \to \pi e\nu$ decay
$f_+(q^2)$ to the point $q^2 =0$, where $q$ is the lepton pair
momentum. Due to the Ademollo-Gatto theorem  \cite{21} the corrections to the CVC
value $f_+(0)=1$ are of the second order of flavor SU(3)
violation, and these small terms were calculated.
(For the case of isotopic SU(2) violation a similar theorem was proved
in \cite{22}). As a
result of this (and other) analyses PDG gives the following
value: $V_{us} \equiv \lambda = 0.2243(5)$.

The accuracy of $\lambda$ is high: the other parameters of CKM
matrix are known much worse.
$V_{cd}$ is measured in the processes with $c$-quark with an order of
magnitude worse accuracy: $V_{cd} = 0.218(4)$.

The value of $V_{cb}$ is determined from the inclusive and
exclusive semileptonic decays of $B$-mesons to charm.
At the level of quarks $b\to cl\nu$ transition is
responsible for these decays: $V_{cb} = (42.2 \pm 0.8)\cdot10^{-3}$.

The value of $|V_{ub}|$ is extracted from the semileptonic
$B$-mesons decays without the charmed particles in the final state
which originated from $b\to ul\nu$ transition:
$V_{ub} = (3.94 \pm 0.36)\cdot10^{-3}$.

The apex of the unitarity triangle should belong to
 a circle on
$(\bar \rho, \bar \eta)$ plane with the center at the point $(0,0)$.
The area between  such two circles (deep green color)  corresponds to the
domain allowed at $2\sigma$.

\subsection{$\varepsilon_K, \Delta m_{B^0}, \Delta m_{B^0_s}$}

CPV in kaon mixing determines the hyperbola shown by light green color
in \Figure\ref{fig:utc}, see Eq.(\ref{89}).

\begin{figure}[h]
  \centering
  \begin{tikzpicture}[scale=1.15]
    \tikzset{every node/.append style={font=\normalsize}}
      \coordinate (A) at (0,0);
      \coordinate (B) at (1,0);
      \coordinate (C) at (3,0);
      \coordinate (D) at (4,0);
      \coordinate (E) at (0,2);
      \coordinate (F) at (1,2);
      \coordinate (G) at (3,2);
      \coordinate (H) at (4,2);
      \coordinate (S1) at (0,1);
      \coordinate (S2) at (4,1);
      \draw[electron] (A) to node[below]{$b$} (B);
      \draw[electron] (B) to node[below]{$u,c,t$} (C);
      \draw[electron] (C) to node[below]{$d$} (D);
      \draw[electron] (F) to node[above]{$ d$} (E);
      \draw[electron] (G) to node[above]{$ u,~ c,~ t$} (F);
      \draw[electron] (H) to node[above]{$ b$} (G);
      \draw[photon] (B) to node[left]{$W$} (F);
      \draw[photon] (C) to node[right]{$W$} (G);
      \draw (S1) node[left]{$B^{0}$};
      \draw (S2) node[right]{$\bar B^{0}$};
      \filldraw (B) circle (1.5pt);
      \filldraw (C) circle (1.5pt);
      \filldraw (F) circle (1.5pt);
      \filldraw (G) circle (1.5pt);
     \end{tikzpicture}
  \caption{$B^0$ -- $\bar B^0$ mixing.}
  \label{fig:bmix}
\end{figure}
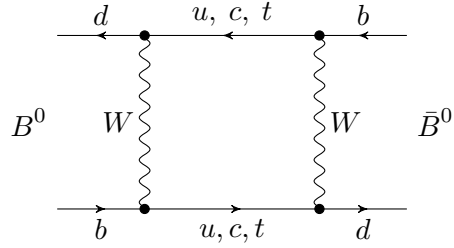

In Standard Model $B_d - \bar B_d$ transition occurs through the
box diagram shown in \Figure\ref{fig:bmix}.
Unlike the case of $K^0 - \bar K^0$
transition the power of $\lambda$ is the same for $u$, $c$ and $t$
quarks inside a loop, so the diagram with $t$-quarks dominates.

Calculating it in complete analogy with $K$-meson case we get:
\begin{equation}
 M_{12} = -\frac{G_F^2
B_{B_d} f_{B_d}^2}{12\pi^2}m_B m_t^2 \eta_B V_{tb}^2 V_{td}^{*^2}
I(\xi) \;\; ,
\label{119}
\end{equation}
where $I(\xi)$ is the same function
as that for $K$-mesons, $\eta_B = 0.55 \pm 0.01$ (NLO).

$\Gamma_{12}$ is determined by the absorptive part of the same diagram
(so, 4 diagrams
altogether: $uu$, $uc$, $cu$, $cc$ quarks in the inner lines).
The result of calculation is:
\begin{equation}
\Gamma_{12} = \frac{G_F^2 B_{B_d} f_{B_d}^2
m_B^3}{8\pi} [V_{cb}V_{cd}^*(1+O(\frac{m_c^2}{m_b^2}))
+V_{ub}V_{ud}^*]^2 \;\; ,
\label{120}
\end{equation}
where the term $O(m_c^2/m_b^2)$ accounts for nonzero $c$-quark
mass.

Using the unitarity of CKM matrix we get:
\begin{equation}
\Gamma_{12} =
\frac{G_F^2 B_{B_d} f_{B_d}^2 m_B^3}{8\pi}[-V_{tb}V_{td}^*
+O(\frac{m_c^2}{m_b^2})V_{cb}V_{cd}^*]^2 \;\; ,
\label{121}
\end{equation}
and the main term in $\Gamma_{12}$ has the same phase as the main
term in $M_{12}$. That is why CPV in mixing of
$B$-mesons is suppressed by an extra factor $(m_c/m_b)^2$ and
is small.
For the difference of masses of the two
eigenstates from
\begin{equation}
M_+ - M_-
-\frac{i}{2}(\Gamma_+ - \Gamma_-) =
2\sqrt{(M_{12} - \frac{i}{2}\Gamma_{12})(M_{12}^* -
\frac{i}{2}\Gamma_{12}^*)}
\label{122}
\end{equation}

we obtain:
\begin{equation}
\Delta m_{B^0} =-\frac{G_F^2 B_{B_d}
f_B^2}{6\pi^2}m_B m_t^2 \eta_B \mid V_{tb}^2 V_{td}^{*^2}\mid
I(\xi),
\label{123}
\end{equation}
and $\Delta m_{B^0}$ is negative as
well as in the kaon system: a heavier state has a smaller width.

\subsection{$\Delta m_{B^0}$ and semileptonic $B^0(\bar B^0)$ decays}

The $B$-meson semileptonic decays are induced by a semileptonic
$b$-quark decay, $b\to c l^- \nu  \;\; (u l^- \nu )$. In this way in the
decays of $\bar B^0$ mesons $l^-$ are produced, while in the
decays of $B^0$ mesons $l^+$ are produced. However, $B^0$ and
$\bar B^0$ are not the mass eigenstates and being produced at
$t=0$ they start to oscillate according to the following formulas:
\begin{equation}
B^0(t) = \frac{e^{-i\lambda_+ t} + e^{-i\lambda_- t}}{2} B^0 +
\frac{q}{p} \frac{e^{-i\lambda_+ t} - e^{-i\lambda_- t}}{2} \bar
B^0 \;\; ,
\label{124}
\end{equation}
\begin{equation}
\bar B^0(t) = \frac{e^{-i\lambda_+
t} + e^{-i\lambda_- t}}{2} \bar B^0 + \frac{p}{q}
\frac{e^{-i\lambda_+ t} - e^{-i\lambda_- t}}{2} B^0 \;\; .
\label{125}
\end{equation}

That is why in their semileptonic decays the ``wrong sign
leptons'' are sometimes produced, $l^-$ in the decays of the
particles born  as $B^0$ and $l^+$ in the decays of the particles
born as $\bar B^0$. The number of these ``wrong sign'' events
depends on the ratio of the oscillation frequency $\Delta m$ and
$B$-meson lifetime $\Gamma$ (unlike the case of $K$-mesons for
$B$-mesons $\Delta\Gamma \ll \Gamma$). For $\Delta
m \gg \Gamma$ a large number of oscillations occurs, and the
number of ``the wrong sign leptons'' equals that of a normal sign.
If $\Delta m \ll \Gamma$, then $B$-mesons decay before they start
to oscillate.

The pioneering detection of ``the wrong sign events''
by ARGUS collaboration  in 1987 demonstrated that $\Delta
m$ is of the order of $\Gamma$, which in the framework of Standard
Model could be understood only if the top quark is unusually heavy,
$m_t \ge 100$ GeV \cite{23}. Fast $B^0 - \bar B^0$ oscillations made
possible the construction of asymmetric $B$-factories (suggested in
\cite{24}) where CPV in
$B^0$ decays was observed. (Let us
mention that UA1 collaboration saw the events which were
interpreted as a possible manifestation of $B_s^0 - \bar B_s^0$
oscillations \cite{25}.)

Integrating the probabilities of $B^0$ decays in $l^+$ and $l^-$
over $t$, we obtain for ``the wrong sign lepton'' probability:
\begin{equation}
W_{B^0 \to \bar B^0} \equiv \frac{N_{B^0 \to l^- X}}{N_{B^0 \to
l^- X} +N_{B^0 \to l^+ X}}
= \frac{\mid \frac{q}{p}\mid^2
(\frac{\Delta m}{\Gamma})^2}{2+(\frac{\Delta m}{\Gamma})^2 +
\mid\frac{q}{p}\mid^2(\frac{\Delta m}{\Gamma})^2} \;\; ,
\label{126}
\end{equation}
where we neglect $\Delta\Gamma$, the difference of
$B_+$- and $B_-$-mesons lifetimes. Precisely according to our
discussion for $\Delta m/\Gamma \gg 1$ we have $W = 1/2$, while
for $\Delta m/\Gamma \ll\ 1$ we have $W = 1/2 (\Delta m/\Gamma)^2$
(with high accuracy $\mid p/q\mid =1$).

For $\bar B^0$ decays we get the same formula with the interchange
of $q$ and $p$.

In ARGUS experiment $B$-mesons were produced in $\Upsilon(4S)$
decays: $\Upsilon(4 S) \to B \bar B$. $\Upsilon$ resonances have
$J^{PC} = 1^{--}$, that is why (pseudo)scalar $B$-mesons are
produced in $P$-wave. It means that $B\bar B$ wave function is
antisymmetric at the interchange  of $B$ and $\bar B$. This fact
forbids the configurations in which due to $B - \bar B$
oscillations both mesons become $B$, or both become $\bar B$.
However, after one of the  $B$-meson decays the flavor of the
remaining one is tagged, and it oscillates according to
(\ref{124}), (\ref{125}).

If the first decay is semileptonic with $l^+$ emission indicating
that a decaying particle was $B^0$, then the second particle was
initially $\bar B^0$. Thus
taking $\mid p/q\mid =1$ we get for the relative number of the
same sign dileptons born in semileptonic decays of $B$-mesons,
produced in $\Upsilon(4S) \to B \bar B$ decays:
\begin{equation}
\frac{N_{l^+
l^+} +N_{l^- l^-}}{N_{l^+ l^-}} =\frac{W}{1-W} = \frac{x^2}{2+
x^2} \; , \;\; x \equiv \frac{\Delta m}{\Gamma} \;\; .
\label{127}
\end{equation}

Let us note that if $B^0$ and $\bar B^0$ are produced incoherently
(say, in hadron collisions) a different formula should be used:
\begin{equation}
\frac{N_{l^+ l^+} +N_{l^- l^-}}{N_{l^+ l^-}} =\frac{2W -
2W^2}{1-2W + 2W^2} = \frac{x^2(2+ x^2)}{2+ 2 x^2 + x^4} \;\; .
\label{128}
\end{equation}

In the absence of oscillations ($x=0$) both equations give
zero; for high frequency oscillations ($x \gg 1$) both of them give
one.

From the time integrated data of ARGUS and CLEO $W_d = 0.182\pm0.015$
follows.
From the
time-dependent analysis  of $B$-decays at the high energy
colliders (LEP II, Tevatron, SLC, LHC) and the time-dependent analysis
at the asymmetric $B$-factories Belle and BaBar the following
result was obtained :
\begin{equation}
x_d = 0.770(4) \;\; .
\label{129}
\end{equation}

By using the life time of $B_d$-mesons:
$\Gamma_{B_d} = [1.52(1) \cdot 10^{-12} \; {\rm sec}]^{-1} \equiv
[1.52(1) {\rm ps}]^{-1}$ we get for the mass difference of $B_d$
mesons:
\begin{equation}
\Delta m_{d} = 0.506(2) {\rm ps}^{-1} \;\; {\rm or, \; equivalently},
W_d = 0.1874\pm 0.0018.
\label{130}
\end{equation}

This $\Delta m_{d}$ value can be used in Eq.(\ref{123})
to extract the value of $|V_{td}|$. The main uncertainty is in a hadronic
matrix element $f_{B_d}\sqrt{B_{B_d}} = 216 \pm 15$ MeV obtained from the lattice QCD
calculations.

\subsection{$\Delta m_{B^0_s}$}

Theoretical uncertainty diminishes in the ratio
\begin{equation}
\frac{\Delta m_s}{\Delta m_{d}} = \frac{m_{B_s}}{m_{B_d}} \xi^2
\frac{|V_{ts}|^2}{|V_{td}|^2},
\label{131}
\end{equation}

where $\xi = (f_{B_s}\sqrt{B_{B_s}})/(f_{B_d}\sqrt{B_{B_d}}) = 1.24 \pm 0.05$.

Since the lifetimes of $B_d$~- and
$B_s$~-mesons are almost equal,  we get:
\begin{equation}
x_{s} \approx x_{d}\frac{|V_{ts}|^2}{|V_{td}|^2}
\label{132}
\end{equation}
which means $x_{s}\gg1$
and very fast
oscillations. That is why $W_{B_s}$ equals $1/2$ with very high
accuracy and one cannot extract $x_{B_s}$ from the time integrated
measurements.

$B^0_s - \bar B^0_s $ oscillations were first observed at Tevatron. The average of
all published measurements
\begin{equation}
\Delta m_{B^0_s} = 17.757 \pm 0.020{\rm (stat)} \pm 0.007
{\rm(syst) \;\; (ps}^{-1})
\label{133}
\end{equation}
is dominated by LHCb.

Thus we get
\begin{equation}
|V_{td}/ V_{ts}| = 0.210\pm0.001({\rm exp})\pm 0.008({\rm theor}),
\label{134}
\end{equation}
which corresponds to yellow (only $\Delta m_{d}$) and brown ($\Delta m_{d}$ and $\Delta m_{s}$) circles in \Figure\ref{fig:utc}.

What remains is the values of the angles of the unitarity triangle, which
are determined by CP-violation measurements in B-meson decays. Soon we will go there.

\subsection{$\Delta \Gamma/\Gamma$}

For the difference of the width of $B_{dL}$ and $B_{dH}$ we
obtain
\begin{equation}
\Delta \Gamma_{B_d} = 2\Gamma_{12}
\approx \frac{G_F^2 B_{B_d} f_B^2 m_B^3}{4\pi} \mid V_{td}\mid^2
\;\; ,
\label{135}
\end{equation}
which is very small:
\begin{equation}
\frac{\Delta\Gamma_{B_d}}{\Gamma_{B_d}} < 1 {\rm \%} \ \;\; ,
\label{136}
\end{equation}
as opposite to $K$-meson case, where $K_S$
and $K_L$ lifetimes differ strongly.

\begin{center}
\begin{figure}[h]
\hspace{25mm}
    \includegraphics[width=.8\textwidth]{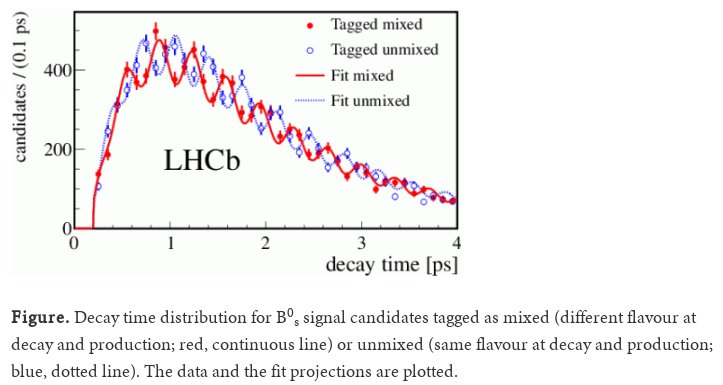}
\caption{$B_s$ -- $\bar B_s$ oscillations \cite{277}.}
\label{fig:bsosc}
\end{figure}
\end{center}

In the $B_s$-meson system a larger time difference was expected;
substituting $V_{ts}$ instead of $V_{td}$  we obtain:
\begin{equation}
\frac{\Delta\Gamma_{B_s}}{\Gamma_{B_s}} \sim 10 {\rm \%} \;\; .
\label{137}
\end{equation}

Here are experimental results:
\begin{equation}
\Gamma_{B^0_{sL}} = (1.414(10){\rm ps})^{-1}
\label{138}
\end{equation}
\begin{equation}
 \Gamma_{B^0_{sH}} = (1.624(14){\rm ps})^{-1},
 \label{139}
 \end{equation}
where  L is light, H - heavy.

\bigskip

\section{CPV in $B^0 - \bar B^0$ mixing}

For a long time CPV in $K$-mesons was observed only in $K^0 - \bar
K^0$ mixing. That is why it seems reasonable to start studying CPV
in $B$-mesons from their mixing:
\begin{eqnarray}
\left |\frac{q}{p}\right | =
\left |\sqrt{1+ \frac{i}{2}\left(\frac{\Gamma_{12}}{M_{12}} -
\frac{\Gamma_{12}^*}{M_{12}^*} \right)}\right | = \left | 1+
\frac{i}{4} \left(\frac{\Gamma_{12}}{M_{12}} -
\frac{\Gamma_{12}^*}{M_{12}^*}\right)\right | = \nonumber \\
=
1-\frac{1}{2} {\rm Im} \left(\frac{\Gamma_{12}}{M_{12}}\right) \approx 1
- \frac{m_c^2}{m_t^2} {\rm Im} \frac{V_{cb} V_{cd}^*}{V_{tb} V_{td}^*}
\approx 1- O(10^{-4}) \;\; .
\label{140}
\end{eqnarray}

We see that CPV in $B_d - \bar B_d$
mixing is very small
because $t$-quark is very heavy and is even smaller in $B_s - \bar
B_s$ mixing.

The experimental observation of $B_d - \bar B_d$
mixing comes from the detection of the same sign leptons produced
in the semileptonic decays of $B_d - \bar B_d$ pair from
$\Upsilon(4S)$ decay. Due to CPV in the mixing the number of $l^-
l^-$ events will differ from that of $l^+ l^+$ and this difference
is proportional to $|\frac{q}{p}| -1  \sim 10^{-4}$:
\begin{equation}
A_{SL}^B = \frac{N (\bar B^0 \to l^+ X) -N (B^0 \to l^- X)}{N(\bar B^0 \to
l^+ X) + N (B^0 \to l^- X)} = O(10^{-4}).
\label{141}
\end{equation}

The experimental number is:
\begin{equation}
 A_{SL}^{B_d} = 0.0021\pm 0.0017 \;\; ,
 \label{142}
 \end{equation}
or
\begin{equation}
|q/p|_{B_d} = 1.0010\pm0.0008 \;\;  .
\label{143}
\end{equation}
This result shows no evidence of CPV and does not constrain the SM.

\bigskip

\section{CPV in interference of mixing and decays,
    $B^0(\bar B^0) \to J/\Psi K$, angle $\beta$.}
\subsection{General formulae.}

 As soon as it became clear that CPV in $B - \bar B$ mixing is
small theoreticians started to look for another way to find CPV in
$B$ decays.
The evident alternative is the direct CPV. It is very
small in $K$-mesons because: a) the third generation almost
decouples in $K$ decays; b) due to $\Delta I = 1/2$ rule. Since in
$B$-meson decays all three quark generations are involved and
there are many different final states, large
direct CPV does occur \cite{301} - \cite{304}. An evident drawback of
this strategy: a branching ratio of $B$-meson decays into any
particular exclusive hadronic mode is very small (just because
there are many modes available), so a large number of $B$-meson
decays are needed. The specially constructed asymmetric $e^+
e^-$-factories Belle (1999-2010) and BaBar (1999-2008)
working at the invariant mass of $\Upsilon(4S)$
discovered CPV in $B^0(\bar B^0)$ decays in 2001 \cite{14}.

The time evolution of the states produced at $t = 0$ as $B^0$ or $\bar
B^0$ is described by (\ref{124}), (\ref{125}).
It is convenient to present these formulae in a little bit
different form:
\begin{equation}
\mid B^0(t)> = e^{-i\frac{M_+ + M_-}{2}t -
\frac{\Gamma t}{2}}\left[\cos(\frac{\Delta mt}{2})\mid B^0 >
+i\frac{q}{p}\sin(\frac{\Delta mt}{2})\mid \bar B^0 >\right]  \;\; ,
\label{144}
\end{equation}
\begin{equation}
\mid\bar B^0(t)> = e^{-i\frac{M_+ + M_-}{2}t - \frac{\Gamma
t}{2}}\left[+i\frac{p}{q}\sin(\frac{\Delta mt}{2})\mid B^0
> + \cos(\frac{\Delta mt}{2})\mid \bar B^0 >\right] \; ,
\label{145}
\end{equation}
where $\Delta m \equiv M_- - M_+ > 0$, and we take $\Gamma_+ =
\Gamma_- = \Gamma$ neglecting their small difference (which should be accounted for
in case of $B_s$).

Let us consider  a decay in some final state $f$. Introducing the
decay amplitudes according to the following definitions:
\begin{equation}
A_f =
A(B^0 \to f) \; , \;\; \bar A_f = A(\bar B_0 \to f) \;\; ,
\label{146}
\end{equation}
\begin{equation}
A_{\bar f} = A(B^0 \to \bar f) \; , \;\; \bar A_{\bar f} = A(\bar
B_0 \to \bar f) \;\; ,
\label{147}
\end{equation}
for the decay probabilities as functions of time we obtain:
\begin{equation}
P_{B^0 \to f}(t) = e^{-\Gamma
t}\mid A_f\mid^2\left[\cos^2(\frac{\Delta mt}{2}) +\left |\frac{q\bar
A_f}{pA_f}\right |^2\sin^2(\frac{\Delta mt}{2})
-{\rm Im}\left(\frac{q\bar A_f}{pA_f}\right)\sin(\Delta mt)\right]  \;\;  ,
\label{148}
\end{equation}
\begin{equation}
P_{\bar B^0 \to \bar f}(t) = e^{-\Gamma t}\mid \bar A_{\bar
f}\mid^2\left[\cos^2(\frac{\Delta mt}{2}) +\left |\frac{p A_{\bar f}
}{q\bar A_{\bar f}}\right |^2\sin^2(\frac{\Delta mt}{2})
-{\rm Im}\left(\frac{p A_{\bar f}}{q\bar A_{\bar f}}\right)\sin(\Delta
mt)\right] \;\; .
\label{149}
\end{equation}
The difference of these two probabilities  signals different types
of CPV: the difference in the first term in brackets appears
due to direct CPV; the difference in the second term -
due to CPV in mixing or due to direct CPV,
and in the last term -- due to CPV in the
interference of $B^0 - \bar B^0$ mixing and decays.

\begin{center}
\begin{figure}[h]
\hspace{25mm}
\includegraphics[width=.6\textwidth]{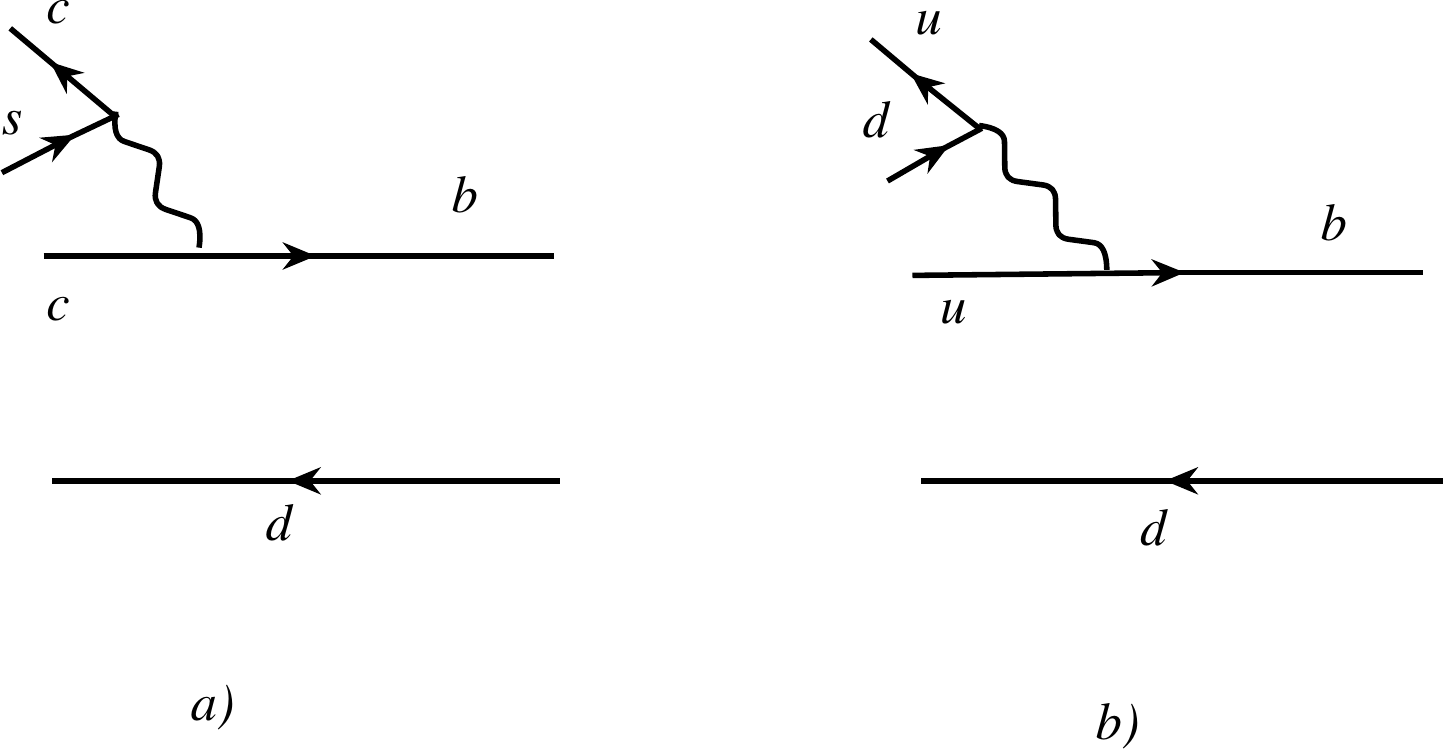}
\caption{Quark diagrams responsible for $B^0 \to J/\Psi K$ and $B^0\to \pi\pi$-decays.}
\label{fig:bpsi}
\end{figure}

  \end{center}

Let $f$ be a CP eigenstate: $\bar f = \eta_f f$, where $\eta_f =
+(-)$ for CP even (odd) $f$. (Two examples of such decays: $B^0
\to J/\Psi K_{S(L)}$ and $B^0 \to \pi^+ \pi^-$ are described by
the quark diagrams shown in \Figure\ref{fig:bpsi}.  The analogous diagrams
describe $\bar B^0$ decays in the same final states.) The
following equalities can be easily obtained:
\begin{equation}
A_{\bar f} = \eta_f A_f
\; , \;\; \bar A_{\bar f} = \eta_f \bar A_f \;\; .
\label{150}
\end{equation}

In the absence of CPV the expressions in brackets  are
equal and the obtained formulas   describe the exponential particle decay
without oscillations. Taking CPV into account and neglecting a
small deviation of $\mid p/q\mid$ from one, for CPV asymmetry of
the decays into CP eigenstate we obtain:
\begin{eqnarray}
a_{CP}(t) \equiv
\frac{P_{\bar B^0\to f} - P_{B^0 \to f}}{P_{\bar B^0 \to f} +
P_{B_0 \to f}} = \frac{\mid\lambda\mid^2 -1}{\mid\lambda\mid^2
+1}\cos(\Delta mt) +  \frac{2 Im\lambda}{\mid\lambda\mid^2 +1}
\sin(\Delta mt) \equiv \nonumber \\
\equiv -C_f \cos(\Delta mt) + S_f
\sin(\Delta mt) \;\; ,
\label{151}
\end{eqnarray}
where $\lambda \equiv
\frac{q \bar A_f}{p A_f}$.  (Do not confuse
with the parameter of CKM matrix).

The nonzero value
of $C_f$ corresponds to direct CPV; it occurs when more than one
amplitude contribute to the decay. For extraction of CPV
parameters (the angles of the unitarity triangle) in this case the
knowledge of strong rescattering phases is necessary. The
nonvanishing $S_f$ describes CPV in the interference of mixing and
decay. It is nonzero even when there is only one decay amplitude,
and $|\lambda | =1$. Such decays are of special interest since the
extraction of CPV parameters becomes independent of poorly known
strong phases of the final particles rescattering.

The decays of $\Upsilon(4S)$ resonance produced in $e^+ e^-$
annihilation  are a powerful source of $B^0 \bar B^0$ pairs. A
semileptonic decay of one of the $B$'s tags ``beauty'' of the
partner at the moment of decay (since $(B^0 B^0), (\bar B^0 \bar B^0)$
states are forbidden) thus making it possible to study
CPV. However, the time-integrated asymmetry is zero for decays were
$C_f$ is zero. This happens since we do not know which of the two
$B$-mesons decays earlier, and asymmetry is proportional to: $ I
= \int\limits^\infty _{-\infty} e^{-\Gamma |t|} \sin (\Delta mt)
dt = 0 \;\; . $
The asymmetric $B$-factories provide possibility to measure the
time-dependence: $\Upsilon(4S)$ moves in a laboratory system, and
since energy release in $\Upsilon(4S) \to B\bar B$ decay is very
small both $B$ and $\bar B$ move with the same velocity as the
original $\Upsilon(4S)$. This makes the resolution of $B$ decay
vertices possible unlike the case of $\Upsilon(4S)$ decay at rest,
when non-relativistic $B$ and $\bar B$ decay at almost the
same point. The implementation of the time-dependent analysis for the search of CPV in $B$-mesons was suggested in \cite{305} - \cite{307}.

\subsection{$\mbox{\boldmath$B_d^0(\bar B_d^0) \to J/\Psi K_{S(L)}$}$,
$\mbox{\boldmath$\sin 2\beta$}$ -- straight lines  }

  The tree diagram contributing to this decay is shown in \Figure\ref{fig:bpsi} a).
The product of the corresponding CKM matrix elements is: $V_{cb}^*
V_{cs} \simeq A\lambda^2$. Also the penguin diagram $b\to sg$ with
the subsequent $g \to c \bar c$ decay contributes to the decay
amplitude. Its contribution is proportional to:
\begin{eqnarray}
P \sim V_{us}
V_{ub}^* f(m_u) + V_{cs} V_{cb}^* f(m_c) + V_{ts} V_{tb}^* f(m_t)
= \nonumber \\
= V_{us} V_{ub}^* (f(m_u) - f(m_t)) + V_{cs} V_{cb}^*
(f(m_c) - f(m_t)) \;\; ,
\label{152}
\end{eqnarray}
where function $f$ describes the contribution of quark loop and we have subtracted zero
from the expression on the first line. The last term on the
second line has the same weak phase as the tree amplitude, while
the first term has a CKM factor $V_{us} V_{ub}^* \sim \lambda^4
(\rho - i\eta)A$. Since (one-loop) penguin amplitude should be in
any case smaller than the tree one, we get that with 1\% accuracy
there is only one weak amplitude governing $B_d^0 (\bar B_d^0) \to
J/\Psi K_{S(L)}$ decays. This is the reason why this mode is
called a ``gold-plated mode'' -- the accuracy of the theoretical
prediction of the CP-asymmetry is very high, and Br $(B_d \to
J/\Psi K^0) \approx 10^{-3}$ is large enough to detect CPV.

\begin{center}
\begin{figure}[h]

    \includegraphics[width=.9\textwidth]{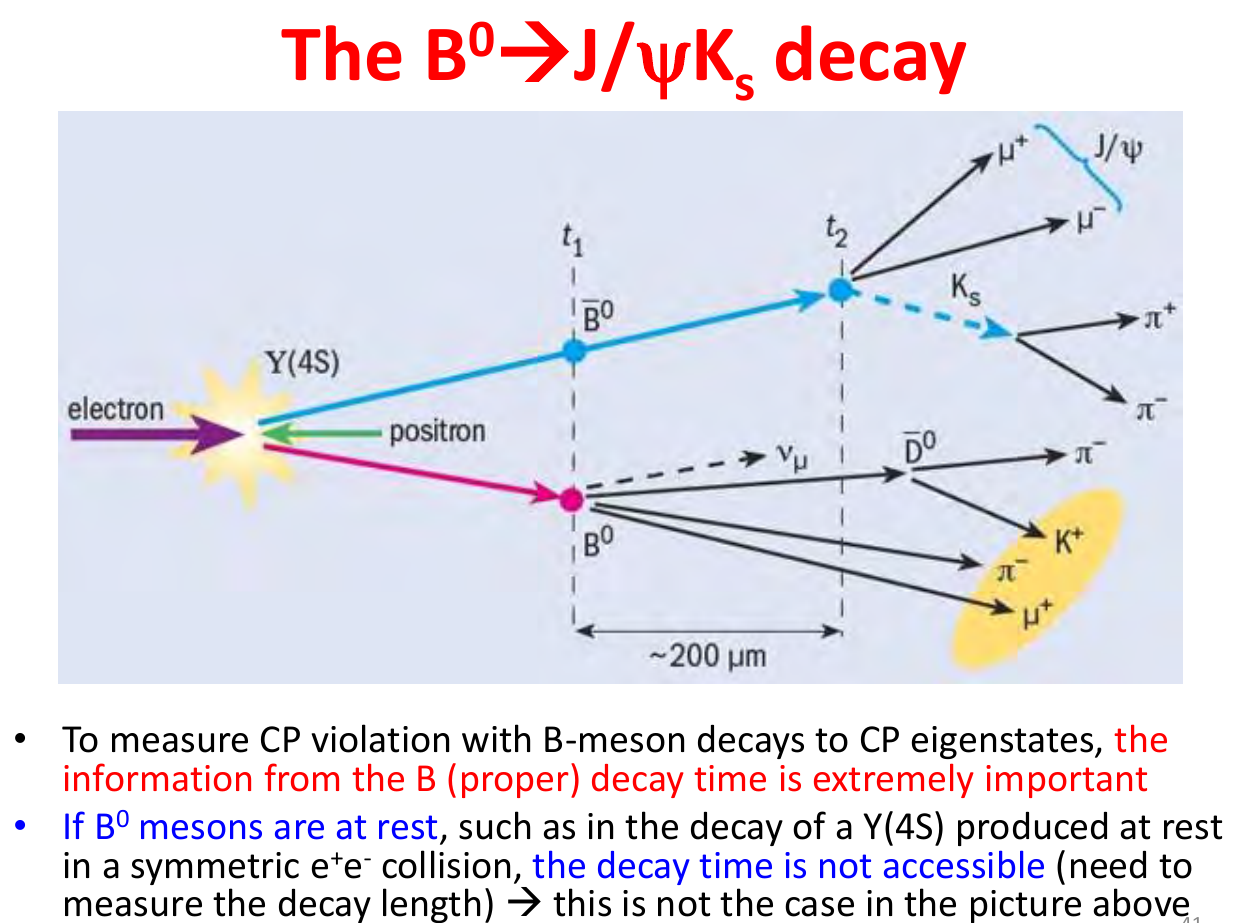}

\caption{Tagging $\bar B^0$-meson by $B^0$-decay.}
\label{fig:bpsik}

    \end{figure}

  \end{center}

Substituting $|\lambda | =1$ in the expression for $a_{CP}(t)$
we obtain:
\begin{equation}
a_{CP}(t) =
{\rm Im} \lambda \sin(\Delta m \Delta t) \;\; , \label{153}
\end{equation}
where
$\Delta t$ is the time difference between the semileptonic decay
of one of $B$-mesons produced in $\Upsilon(4S)$ decay and that of
the second one to $J/\Psi K_{S(L)}$. Using the following equation
\begin{equation}
\bar A_f = \eta_f \bar A_{\bar f} \;\; ,
\label{154}
\end{equation}
where $\eta_f$ is CP parity of the final state, we obtain:
\begin{equation}
\lambda =
\left(\frac{q}{p}\right)_{B_d} \frac{A_{\bar B^0 \to J/\Psi
K_{S(L)}}}{A_{B^0 \to J/\Psi K_{S(L)}}} =
\left(\frac{q}{p}\right)_{B_d} \eta_f \frac{A_{\bar B^0 \to
\overline{J/\Psi K_{S(L)}}}}{A_{B^0 \to J/\Psi K_{S(L)}}} \;\; .
\label{155}
\end{equation}

The amplitude in the nominator contains $\bar K^0$ production. To
project it on $\bar K_{S(L)}$ we should use:
\begin{equation}
\overline{K^0} =
\frac{K_S - K_L}{(q)_K} = \frac{\bar K_S + \bar K_L}{(q)_K} \;\; ,
\label{156}
\end{equation}
getting $(q)_K$ in the denominator. The amplitude in
the denominator contains $K^0$ production, and using:
\begin{equation}
K^0 =
\frac{K_S + K_L}{(p)_K}
\label{157}
\end{equation}
we obtain factor
$(p)_K$ in the nominator. Collecting all the factors together and
substituting CKM matrix elements for $\bar A_{\bar f}/A_f$ ratio
we get:
\begin{equation}
\lambda = \eta_{S(L)} \left(\frac{q}{p}\right)_{B_d}
\frac{V_{cb} V_{cs}^*}{V_{cb}^* V_{cs}} \left(\frac{p}{q}\right)_K
\;\; .
\label{158}
\end{equation}

\begin{center}
\begin{figure}[h]

    \includegraphics[width=.9\textwidth]{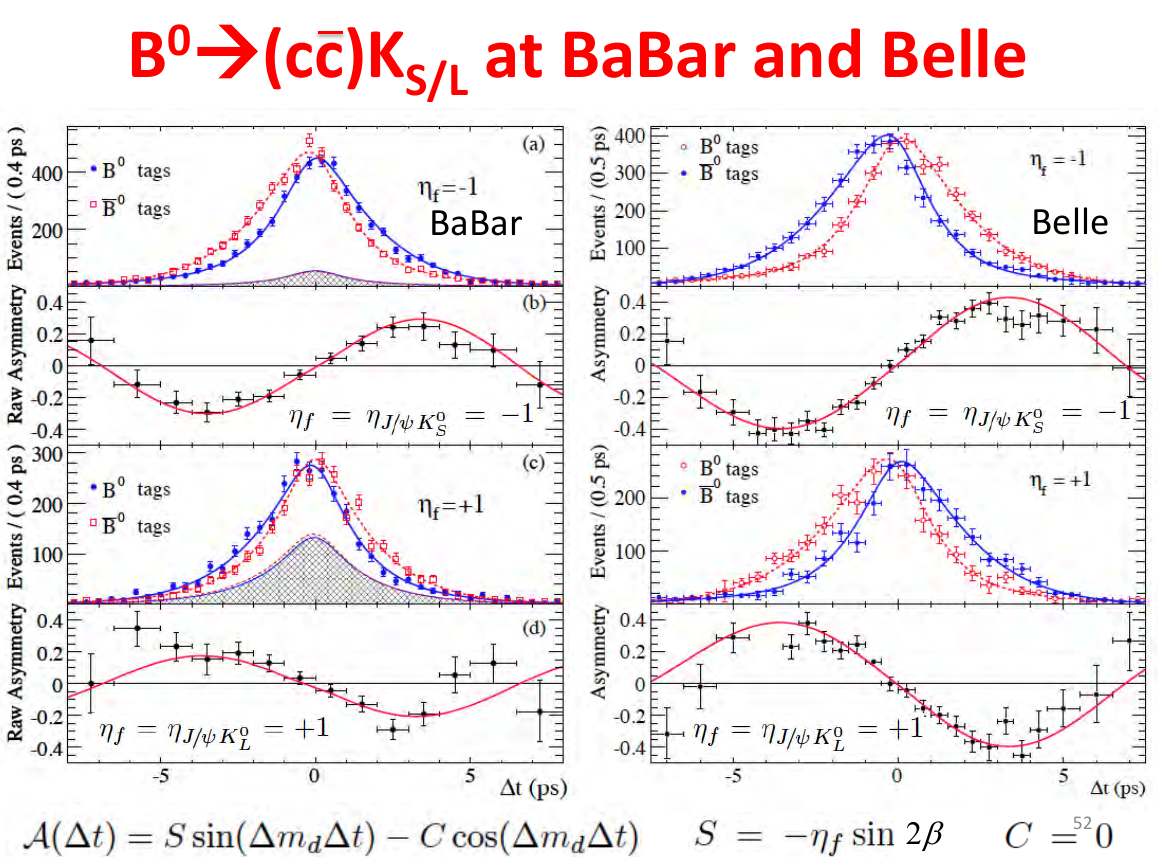}
\caption{Measurements of CPV asymmetries.}
\label{fig:bacp}
\end{figure}
\end{center}

Substituting the expressions for $(q/p)_{B_d}$ and $(p/q)_K$
we obtain:
\begin{equation}
\lambda(J/\Psi K_{S(L)}) = \eta_{S(L)}\frac{V_{td} V_{tb}^*}{V_{td}^*
V_{tb}} \frac{V_{cb} V_{cs}^*}{V_{cb}^* V_{cs}} \frac{V_{cd}^*
V_{cs}}{V_{cd} V_{cs}^*} \;\; ,
\label{159}
\end{equation}
which is invariant
under the phase rotation of any quark field. From the unitarity
triangle figure we have
\begin{equation} \arg (V_{tb}^* V_{td}) = \pi - \beta \;\; ,
\label{160}
\end{equation}
 and we finally obtain:
 \begin{equation}
 a_{CP}(t)\left
|_{_{_{_{\large J/\Psi K_{S(L)}}}}}  = - \eta_{S(L)}\sin(2\beta)
\sin(\Delta m \Delta t) \right. \;\; ,
\label{161}
\end{equation}
which is a simple prediction of the Standard Model.
Since in $B$ decays $J/\Psi$ and $K_{S(L)}$ are produced in
$P$-wave, $\eta_{S(L)} = -1(+1)$ (CP of $J/\Psi$ is $+1$, that of
$K_S$ is $+1$ as well, and $(-1)^l = -1$ comes from $P$-wave; CP of
$K_L$ is $-1$).

In this way the measurement
of this asymmetry at $B$-factories provides the value of  angle
$\beta$ of the unitarity triangle.
The  Belle, BaBar  and LHCb average  is:
\begin{equation}
\sin 2\beta = 0.691 \pm 0.017 \;\; ,
\label{162}
\end{equation}
which corresponds to
\begin{equation}
\beta = (21.9\pm0.7)^0 .
\label{163}
\end{equation}
As a final state not only $J/\Psi
K_{S(L)}$ were selected, but neutral kaons with the other charmonium
states as well.

Let us note that the decay amplitudes
and $K^0 - \bar K^0$ mixing do not contain a complex phase, that is
why the only source of it in $B^0 \to {\rm
charmonium} \; K_{S(L)}$
decays is $B^0 - \bar B^0$ mixing:
\begin{equation}
\left(\frac{q}{p}\right)_{B_d} = \sqrt\frac{M_{12}^*}{M_{12}} =
\frac{V_{tb}^* V_{td}}{V_{tb} V_{td}^*} \;\; ,
\label{164}
\end{equation}
thus the phase comes from $V_{td}$, that is why the final
expression  contains  angle $2\beta$ -- the phase of
$V_{td}/V_{td}^*$.

\Figure\ref{fig:bpsik} and \Figure\ref{fig:bacp} (see \cite{aaa}) illustrate the above discussion.

\bigskip

\section{What is the probability of $ \Upsilon(4S) \rightarrow B_d^0\bar B_d^0
\to J/\Psi K_{S}  \;J/\Psi K_{S}$ decay?}

The following parameters are used to describe the time evolution of $B$-mesons:
$
m\equiv (m_H + m_L)/2 \; , \;\; \Delta m \equiv m_H - m_L \; , \;\; \Gamma_H = \Gamma_L = \Gamma \; .
$

Since $J^{PC} (\Upsilon)= 1^{--}$, $B$-mesons are produced in P-wave, so their
wave function is $C$-odd:
$
\Psi(t_1, t_2) = B^0(t_1) \bar B^0 (t_2) -  B^0(t_2) \bar B^0(t_1).$

For the decay amplitude we get:
\begin{eqnarray}
\left< J/\Psi K_{S} \; J/\Psi K_S| \Psi(t_1, t_2)\right> = e^{-im t_1 - \frac{\Gamma t_1}{2}}\left[ A\cos \frac{\Delta m t_1}{2} +  i\frac{q}{p}  \sin\left(\frac{\Delta mt_1}{2}\right) \bar A \right] \times \nonumber \\
\times e^{-im t_2 - \frac{\Gamma t_2}{2}}\left[ \cos \left(\frac{\Delta m t_2}{2}\right) \bar A +
i\frac{p}{q} \sin\left(\frac{\Delta m t_2}{2}\right) A \right] - (t_1 \leftrightarrow t_2) = ~~~ \\
= e^{-im(t_1 + t_2) - \Gamma\frac{t_1 + t_2}{2}} \left[(i\frac{p}{q} A^2 - i\frac{q}{p} \bar A^2) \cos \left(\frac{\Delta m t_1}{2}\right) \sin\left(\frac{\Delta m t_2}{2}\right)\right. + \nonumber \\
+\left.(i\frac{q}{p} \bar A^2 - i\frac{p}{q} A^2) \sin\left(\frac{\Delta m t_1}{2}\right)\cos\left(\frac{\Delta m t_2}{2}\right)\right]
= -e^{-2imt-\Gamma t}(i\frac{p}{q}A^2)[1-\lambda^2] \sin \left(\frac{\Delta m\Delta t}{2}\right) \;\; , \nonumber
\label{165}
\end{eqnarray}
where $t\equiv \frac{t_1 + t_2}{2} \; ,  \Delta t \equiv t_1-t_2,  \;\; \frac{q}{p} = e^{-2i\beta}.$

Decay probability equals
\begin{equation}
P(J/\Psi K_S, J/\Psi K_S) = e^{-2\Gamma t} |A|^4 [1-e^{4i\beta}][1-e^{-4i\beta}] \sin^2 \left(\frac{\Delta m \Delta t}{2}\right) \sim e^{-2\Gamma t} \sin^2(2\beta) \sin^2 \frac{(\Delta m \Delta t)}{2} \; .
\label{166}
\end{equation}

Changing integration variables in the expression for decay probability according to
\begin{equation}
\int\limits_0^\infty dt_1 \int\limits_0^\infty dt_2 = \int\limits_{-\infty}^{\infty} d(\Delta t) \int\limits_{|\Delta t|/2}^\infty dt
\label{167}
\end{equation}

and performing integration over $t$ we get:
\begin{equation}
N(\Delta t) \sim \sin^2 2\beta [1-\cos(\Delta m \Delta t)] e^{-\Gamma|\Delta t|} \;\; ,
\label{168}
\end{equation}
which is zero when $\Delta t = 0$ due to Bose statistics, when $\Delta m = 0$ -- no oscillations,
and for $\beta = 0$ -- no CPV (CP $\Upsilon = +$,
CP $(J/\Psi K_S \; J/\Psi K_S) = -$).

For the total number of
$ \Upsilon(4S) \to J/\Psi K_{S}  \;J/\Psi K_{S}$ decays integrating over $\Delta t$ we obtain:
\begin{equation}
N(J/\Psi K_S \; J/\Psi K_S) \sim \sin^2 2\beta \left(\frac{\Delta m^2}{\Delta m^2 + \Gamma^2}\right)
\label{169}
\end{equation}

After one of $B$ decays to $J/\Psi K_{S}$ the second one starts to
oscillate and may decay to $J/\Psi K_{S}$ as well. The initial state is $CP$ even, the final
state is $CP$ odd, so no decays without CPV would occur.

Taking different initial and final states one may
solve many problems
the same way as we have just shown.

$C$-even initial state:
\begin{equation}
\Psi(t_1, t_2) = B^0(t_1) \bar B^0 (t_2) + B^0(t_2) \bar B^0(t_1) \;\; .
\label{170}
\end{equation}
The "classical`` initial state (produced in hadron collisions):
\begin{equation}
\Psi(t_1, t_2) = B^0(t_1) \bar B^0 (t_2) \;\; .
\label{171}
\end{equation}

\section{CPV in $b \to s g \to s s
\bar{s}$ transition: penguin domination}

\begin{figure}[h]
  \centering
  \begin{tikzpicture}[scale=0.8]
 \tikzset{every node/.append style={font=\normalsize}}
    %
    %
    \coordinate (A) at (0,0);
    \coordinate (D) at (4,0);
    \coordinate (E) at (0,2);
    \coordinate (F) at (1,2);
    \coordinate (G) at (3,2);
    \coordinate (H) at (4,2);
    \coordinate (S1) at (0,1);
    \coordinate (S2) at (4,1);
    \coordinate (S3) at (4,3.5);
    \coordinate (C2) at (2,2);
    \coordinate (X) at (2.5,3.5);
    \coordinate (Y) at (4,4.2);
    \coordinate (Z) at (4,2.8);
    \draw[electron] (A) node[above]{$d$} to (D) node[above]{$d$};
    \draw[electron] (F) to (E) node[below]{$b$};
    \draw[electron] (C2) to node[above]{$u,t,$} (F);
    \draw[electron] (G) to node[above]{$c$} (C2);
    \draw[electron] (H) node[below]{$s$} to (G);
    \draw[electron] (X) to (Y) node[right]{$s$};
    \draw[electron] (Z) node[right]{$s$} to (X);
    \draw[photon] (C2) to node[left]{$g$} (X);
    \draw[photon] (F) ++ (0,-1.6pt) arc (188:360:1 and 0.76) ;
    \filldraw (X) circle (1.5pt);
    \filldraw (C2) circle (1.5pt);
    \filldraw (F) circle (1.5pt);
    \filldraw (G) circle (1.5pt);
  \end{tikzpicture}
 \caption{Penguin diagram describing $b \to ss \bar s$-transition.}
\label{fig:bs}
\end{figure}
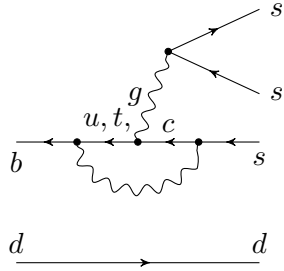

The decays
$ B_d\to \phi K^0, K^+ K^- K^0, \eta^\prime K^0$ proceed through the diagrams shown in \Figure\ref{fig:bs}.

The diagram with an intermediate $u$-quark is
proportional to $\lambda^4$, while those with intermediate $c$-
and $t$-quarks are proportional to $\lambda^2$. In this way the
main part of the decay amplitude is free of CKM phase, just like
in case of $B_d \to J/\Psi K$ decays. A nonzero phase which leads
to time-dependent CP asymmetry comes from $B_d - \bar B_d$
transition:
\begin{equation}
a_{CP}(t) = -\eta_f \sin(2\beta) \sin(\Delta m
\Delta t) \;\; ,
\label{172}
\end{equation}
analogously to $B_d \to J/\Psi K$ decays.

\begin{center}
\begin{figure}[h]
\hspace{30mm}
    \includegraphics[width=.5\textwidth]{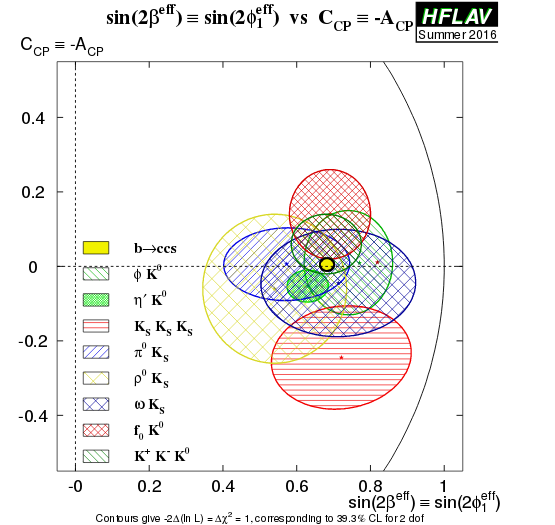}
\caption{CP-asymmetries from $B_d$-decays with production of three strange quarks.}
\label{fig:bsss}
\end{figure}

\end{center}

The main interest in these decays is to look for phases of NP which may be hidden in loops. According to \Figure\ref{fig:bsss} \cite{32} SM nicely describes the experimental data within their present day accuracy.

\section{$\mbox{\boldmath$B_s(\bar B_s) \to J/\Psi \phi$}$,
$\mbox{\boldmath$\phi_s$}$}

This decay is an analog of $B^0(\bar B^0) \to J/\Psi K$ decay: the tree amplitude
dominates and CP asymmetry could appear from
$B_s\leftrightarrow\bar B_s$ transition. $V_{ts}$ unlike $V_{td}$ is
almost real, so asymmetry should be very small in SM -- a good
  place to look for New Physics. The angular analysis of
$J/\Psi \rightarrow \mu^+\mu^-$ and $\phi\rightarrow K K$ decays is
necessary to select the final states with definite CP parity.

Taking the difference of the width of two eigenstates into account
($\Delta\Gamma = \Gamma_L - \Gamma_H$) we get:
\begin{equation}
P_{B_s\to f}(t)=\frac{1}{2}e^{-\Gamma t}|A_f|^2(1+|\lambda_f|^2)[\cosh(\Delta\Gamma t/2)-D_f\sinh (\Delta\Gamma t/2)+  C_f\cos(\Delta m t)- S_f\sin(\Delta m t)] \;\; ,
\label{173}
\end{equation}
\begin{equation}
P_{\bar B_s\to f}(t)=\frac{1}{2}e^{-\Gamma t}|\frac{p}{q}A_f|^2(1+|\lambda_f|^2)
[\cosh(\Delta\Gamma t/2)-D_f\sinh(\Delta\Gamma t/2)- C_f\cos(\Delta m t)+ S_f\sin(\Delta m t)] \;\; ,
\label{174}
\end{equation}
\begin{equation}
D_f=\frac{2Re\lambda_f}{1+|\lambda_f|^2},\;\;\; C_f=\frac{1-|\lambda_f|^2}{1+|\lambda_f|^2},
\;\;\;S_f=\frac{2Im\lambda_f}{1+|\lambda_f|^2} \;\; .
\label{175}
\end{equation}
\begin{equation}
A_{CP}(t)(|p/q|=1) =\frac{-C_f\cos(\Delta m t) + S_f\sin(\Delta m t)}
{\cosh(\Delta\Gamma t/2) - D_f\sinh(\Delta\Gamma t/2)} \;\; .
\label{176}
\end{equation}

Standard Model prediction is $\phi_s^{SM} = -$arg$\frac{V_{ts}V_{tb}^*}{V_{ts}^*V_{tb}} =
-2\lambda^2\eta$ =-0.036 rad,
while $\phi_s^{exp} = -0.040\pm0.025$ rad. No New Physics in this decay as well.

\bigskip

\section{Angles $\alpha$ and $\gamma$}

\subsection{$\alpha: B \longrightarrow \pi\pi, \rho\rho, \pi\rho $}
Since $\alpha$ is the angle between $V_{tb}^*V_{td}$
and $V_{ub}^*V_{ud}$, the time dependent $CP$ asymmetries
in $b\longrightarrow u \bar u d$ decay dominated modes
directly measure $\rm{sin}(2\alpha)$.

$b\longrightarrow d$ penguin amplitudes have different CKM phases compared to
the tree amplitude and are of the same order in $\lambda$. Thus the
penguin contribution can be sizeable, making
determination of $\alpha$ complicated.

Fortunately $Br(B\rightarrow\rho^0\rho^0) \ll
Br(B\rightarrow\rho^+\rho^-), Br(B^+\rightarrow\rho^+\rho^0) $,
which proves that the contribution of the penguins in
 $B \longrightarrow \rho\rho$ decays is small.

Moreover, the longitudinal polarization fractions in
$B\rightarrow\rho^+\rho^-, B^+\rightarrow\rho^+\rho^0$
decays appeared to be close to unity, which means that
the final states are CP even and the following relations
should be valid:
\begin{equation}
S_{\rho^+\rho^-} = {\rm sin}(2\alpha),\;\; \;\;C_{\rho^+\rho^-} = 0 \;\; .
\label{177}
\end{equation}

The experimental numbers are:
\begin{equation}
S_{\rho^+\rho^-} = -0.05\pm0.17,\;\; \;\;C_{\rho^+\rho^-} = -0.06\pm0.13 \;\; .
\label{178}
\end{equation}

So, $C$ is compatible with zero,
while from $S$ we get
\begin{equation}
\alpha = (91\pm5)^0 \;\; .
\label{179}
\end{equation}
Finally from the combination of the
$B \longrightarrow \pi\pi, \rho\rho, \pi\rho $ modes the following result
is obtained:   $\alpha = (85\pm4)^0.$

\bigskip
Problem 8

In the decays considered in this section the quarks of the first and the third generations participate, so only 2 generations are     involved. As it has been stated and demonstrated, at least 3 generations are needed for CPV. So, how does it happen that in $B\longrightarrow\rho\rho$ decays CP is violated?

\subsection{$\gamma$}

The next task is to measure angle $\gamma$, or the phase of $V_{ub}$. In $B_d$ decays
angle $\beta$ enters the game through $B_d - \bar B_d$ mixing. To
avoid it in order to single out angle $\gamma$ we should consider
$B_s$ decays, or the decays of charged $B$-mesons \cite{320}.
The interference of $B^- \longrightarrow D^0 K^- (b \longrightarrow c \bar u
s)$ and $B^- \longrightarrow \bar D^0 K^- (b\longrightarrow u \bar c
s)$ transitions in the final states accessible in both $D^0$
 and $\bar D^0$ decays (such as $K^0_S \pi^+\pi^-$) provides the best accuracy
 in $\gamma$ determination \cite{26}. Combining all the existing methods, the
 following result was obtained:
\begin{equation}
\gamma = (74\pm5)^0 \;\; .
\label{180}
\end{equation}

Here LHCb measurement is significantly more precise than old Belle and BaBar results and it undergoes continuous improvement.

\section{CKM fit}

UTfit and CKMfitter collaborations are making fits of available data
by four Wolfenstein parameters. Here are UTfit results:
\begin{eqnarray}
\lambda &=& 0.225(1) \;\; , \nonumber \\
A &=& 0.83(1) \;\; , \nonumber \\
\eta &=& 0.36(1) \;\; , \nonumber \\
\rho &=& 0.15(1) \;\; .
\label{181}
\end{eqnarray}

For the angles of unitarity triangle the result of fit is:
\begin{equation}
\alpha = (90\pm2)^0, \;\;\;\beta = (24\pm1)^0,\;\;\; \gamma =
(66\pm2)^0 \;\; .
\label{182}
\end{equation}

So $\alpha + \beta + \gamma =180^0$ -- no traces of New Physics yet.

The quality of fit is high and CKMfitter results are approximately the same.

\bigskip

\section{Perspectives: $K\longrightarrow\pi\nu\nu$,  Belle II, LHC}

Two running experiments are measuring the probabilities of $K^+ \to \pi^+ \nu\bar\nu$ (NA62 at SPS, CERN) and $K_L \to \pi^0 \nu\bar\nu$ (KOTO at $J$-PARC, Japan) decays. These decays are very rare. In the framework of the SM branching ratios of these decays are predicted with high accuracy: ${\rm Br}(K^+ \to \pi^+ \nu\bar\nu) = (8.4 \pm 1) 10^{-11}$,
${\rm Br}(K_L \to \pi^0 \nu\bar\nu) = (3.4 \pm 0.6) 10^{-11}$. Smallness of branching ratios in the SM makes these decays a proper place to look for indirect manifestations of New Physics.

Belle II experiment at KEK laboratory started taking data in 2019. With much higher luminosity than that collected by Belle and BaBar it will also contribute to the search for New Physics. The planned Belle II sensitivities for the measurement of the angles of unitarity triangle are  1\%.

The knowledge of the unitarity triangle parameters with better accuracy is expected from the future LHC data. Assuming a reasonable improvement of nonperturbative quantities from lattice QCD we can hope that it will be sufficient to crack the triangle.

The useful introductions to flavor physics and CP violation can be found in \cite{27} - \cite{30}

 \subsection*{Acknowledgements}

I am grateful to the organizers of the PINP 2019 Winter School and European School of High Energy Physics 2019 for the invitations to lecture at the Schools, to A.E. Bondar, V.F. Obraztsov and P.N. Pakhlov for useful comments and to S.I. Godunov and E.A. Ilyina for the help in preparation slides and lectures. This work was supported by RSF grant 19-12-00123.


\begin{thebibliography}{999}

\bibitem{0}
S.L. Glashow, {\em Nucl. Phys.} {\bf 22} (1961) 579; \\
S. Weinberg, {\em Phys. Rev. Lett.} {\bf 19} (1967) 1264; \\
A. Salam, {\em Elementary Particle Theory} (Ed. N. Svartholm. - Almquist and Wiksell, 1968, p. 367).
\bibitem{1}
N. Cabibbo, {\em Phys. Rev. Lett.} {\bf 10} (1963) 531.
\bibitem{2}
M. Gell-Mann, {\em Phys. Rev.} {\bf 125} (1962) 1067.
\bibitem{3}
S.L. Glashow, J. Iliopoulos, L. Maiani, {\em Phys. Rev. D} {\bf 2} (1970) 1285.
\bibitem{4}
M. Kobayashi, T. Maskawa, {\em Progr. Theor. Phys.} {\bf 49} (1973) 652.
\bibitem{5}
L. Wolfenstein, {\em Phys. Rev. Lett.} {\bf 51} (1983) 1945.
\bibitem{6}
C. Jarlscog, {\em Phys. Rev. Lett.} {\bf 55} (1985) 1039.
\bibitem{7}
T.D. Lee, C.N. Yang, {\em Phys. Rev.} {\bf 104} (1956) 254.
\bibitem{8}
B.L. Ioffe, L.B. Okun, A.P. Rudik, {\em ZhETF} {\bf 32} 396.
\bibitem{9}
M. Gell-Mann, A. Pais, {\em Phys. Rev.} {\bf 97} (1955) 1387.
\bibitem{888}
T.D. Lee. C.N Yang, R. Oehme, {\em Phys. Rev.} {\bf 106} (1957) 340.
\bibitem{10}
L.D. Landau, {\em ZhETF} {\bf 32} (1957) 405; \\
L.D. Landau, {\em Nucl. Phys.} {\bf 3} (1957) 127.
\bibitem{11}
L.B. Okun, {\em Slaboe vzaimodeistvie elementarnykh chastits} (M.: Fizmatgiz, 1963) (in Russian).
\bibitem{12}
J.H. Christenson, J.W. Cronin, V.L. Fitch, R. Turlay, {\em Phys. Rev.} {\bf 13} (1964) 138.
\bibitem{13}
V. Fanti {\em et. al.} (NA 48 Collaboration), {\em Phys. Rev. Lett.} {\bf 13} (1999) 355; \\
A. Alavi-Harati {\em et al.} (K TeV Collaboration), {\em Phys. Rev. Lett.} {\bf 83} (1999) 22.
\bibitem{14}
B. Aubert {\em et. al.} (BaBar Collaboration), {\em Phys. Rev. Lett.} {\bf 87} (2001) 091801; \\
K. Abe {\em et al.} (Belle Collaboration), {\em Phys. Rev. Lett.} {\bf 87} (2001) 091802.
\bibitem{15}
R.Aaij {\em et. al.} (LHCb Collaboration), {\em Phys. Rev. Lett.} {\bf 122} (2019) 211803.
\bibitem{16}
A.D. Sakharov, {\em Pisma v ZhETF} {\bf 5} (1967) 32; \\
A.D. Sakharov, {\em JETP Lett.} {\bf 5} (1967) 24.
\bibitem{17}
M.I. Vysotsky, {\em Yad. Fiz.} {\bf 31} (1980) 1535; \\
M.I. Vysotsky, {\em Sov. J. Nucl. Phys.} {\bf 31} (1980) 797.
\bibitem{18}
R.Aaij {\em et. al.} (LHCb Collaboration), {\em Phys. Rev. Lett.} {\bf 110} (2013) 221601.
\bibitem{19}
M. Tanabashi {\em et al.}, (Particle Data Group), {\em Phys. Rev. D} {\bf 98} (2018) 030001.
\bibitem{20}
B. Pontecorvo, {\em ZhETF} {\bf 34} (1957) 247; \\
B. Pontecorvo, {\em Sov. Phys. JETP} {\bf 7} (1958) 172.
\bibitem{31}
M. Tanabashi {\em et al.}, (Particle Data Group), {\em Phys. Rev. D} {\bf 98} (2018) 030001; A.Ceccucci, Z. Ligeti, Y. Sakai, CKM Quark-Mixing Matrix.
\bibitem{21}
M. Ademollo, R. Gatto, {\em Phys. Rev. Lett.} {\bf 13} (1964) 264.
\bibitem{22}
M.V. Terent'ev, {\em Sov. Phys. JETP} {\bf 17} (1963) 890.
\bibitem{23}
H. Albrecht {\em et. al.}, {\em Phys. Lett. B} {\bf 192} (1987) 245.
\bibitem{24}
P. Oddone, Proceedings of the Conference on Linear Collider B-B Factory Conceptual design, Los Angeles, California, 26-30 Jan. 1987, edited by D.H. Stork (New Jersey: World Scientific, 1987).
\bibitem{25}
C. Albajar {\em et al.}, {\em Phys. Lett. B} {\bf 186} (1987) 247.
\bibitem{277}
M. Tanabashi {\em et al.}, (Particle Data Group), {\em Phys. Rev. D} {\bf 98} (2018) 030001; O. Schneider, $B^0$ -- $\bar B^0$ mixing.
\bibitem{301}
A.A. Anselm, Ya. I. Azimov, {\em Phys. Lett. B}  {\bf 85} (1979) 72.
\bibitem{302}
M. Bander, D. Silverman, A. Soni, {\em Phys. Rev. Lett.} {\bf 43} (1979) 242.
\bibitem{303}
A. Carter, A. Sanda, {\em Phys. Rev. Lett.} (1980) 952.
\bibitem{304}
I.I. Bigi, A.I. Sanda, {\em Nucl. Phys. B} {\bf 193} (1981) 85.
\bibitem{305}
I. Dunietz, J. Rosner, {\em Phys. Rev. D} {\bf 34} (1986) 1404.
\bibitem{306}
Ya. I. Azimov, N.G. Uraltzev, V.A. Khoze, {\em Yad. Fiz.} {\bf 45} (1987) 1412; \\
Ya. I. Azimov, N.G. Uraltzev, V.A. Khoze, {\em Sov. J. Nucl. Phys.} {\bf 45} (1987) 878.
\bibitem{307}
Ya. I. Azimov, N.G. Uraltzev, V.A. Khoze, {\em Proceedings of XXI LIYaF Winter School} (LIYaF, 1986) p. 178 ({\em in Russian}).
\bibitem{aaa}
Ed. A.J. Bevan {\em et al.}, {\em Eur. Phys. J. C} {\bf 74} (2014) 3026.
\bibitem{32}
M. Tanabashi {\em et al.}, (Particle Data Group), {\em Phys. Rev. D} {\bf 98} (2018) 030001; T. Gershon, Y. Nir, CP Violation in the Quark Sector.
\bibitem{320}
M. Gronau, D. Wyler, {\em Phys. Lett. B} {\bf 265} (1991) 172; \\
M. Gronau, D. London, {\em Phys. Lett. B} {\bf 253} (1991) 483;\\
D. Atwood, I. Dunietz, A. Soni, {\em Phys. Rev. Lett.} {\bf 78} (1997) 3257.
\bibitem{26}
A. Bondar, Proceedings of BINP Special Meeting on Dalitz Analysis, 24-26 Sep. 2002 (unpublished); \\
A. Giri {\em et al.}, {\em Phys. Rev. D} {\bf 68} (2003) 054018.
\bibitem{27}
J. Zupan, {\em arXiv:1903.05062}.
\bibitem{28}
M. Blanke, {\em arXiv:1704.03753}.
\bibitem{29}
B. Grinstein, {\em arXiv:1701.06916}.
\bibitem{30}
S. Gori, {\em arXiv:1610.02629}.


\end{thebibliography}
\end{document}